\documentclass{article}
\usepackage{kbf_preprint,times}

\usepackage{amsmath,amsfonts,bm}

\def\eqref#1{equation~\ref{#1}}

\def\1{\bm{1}}

\DeclareMathAlphabet{\mathsfit}{\encodingdefault}{\sfdefault}{m}{sl}
\SetMathAlphabet{\mathsfit}{bold}{\encodingdefault}{\sfdefault}{bx}{n}

\usepackage{xcolor}
\usepackage{hyperref}
\usepackage{url}

\usepackage{amsthm}
\usepackage{booktabs}

\usepackage{microtype}
\usepackage{wrapfig}
\usepackage{needspace}
\usepackage{listings}

\usepackage[ruled,vlined,linesnumbered]{algorithm2e}

\usepackage{soul}

\usepackage{tabularx}
\usepackage{multirow}
\usepackage{graphicx}
\usepackage{enumerate}
\usepackage{enumitem}
\usepackage{amsmath}
\usepackage{aliascnt,lscape}

\usepackage{amsfonts}
\usepackage{mdframed}
\usepackage{caption}
\usepackage{subcaption}

\usepackage{multicol}
\usepackage{calc}
\usepackage{cleveref}

\usepackage{float}
\usepackage{placeins}
\usepackage{thm-restate}
\usepackage{xstring}
\usepackage{pifont}
\usepackage{xspace}
\usepackage{orcidlink}

\newcommand{\name}{\textsc{KBF}\xspace}

\UseRawInputEncoding

\newcommand{\ignore}[1]{}
\newcommand{\inlineheading}[1]{\noindent\textbf{#1}\hspace{0.5em}}
\newcommand{\partitle}[1]{\inlineheading{#1.}}

\newcommand{\Description}[1]{}

\let\cite\citep

\title{KBF: Knowledge Boundary as Fingerprint for Language Model and Black-Box API Auditing}

\author{%
Yijia Fang\textsuperscript{1}\quad
Yiqing Feng\textsuperscript{2}\quad
Bingyu Li\textsuperscript{3}\quad
Mingxun Zhou\textsuperscript{1}\thanks{Corresponding author.}\\[6pt]
{\normalsize\textsuperscript{1}HKUST, Hong Kong SAR, China}\\
{\normalsize\textsuperscript{2}Xidian University, China}\\
{\normalsize\textsuperscript{3}Beihang University, China}}
\date{}
\hypersetup{
    pdftitle={KBF: Knowledge Boundary as Fingerprint for Language Model and Black-Box API Auditing},
    pdfauthor={Yijia Fang, Yiqing Feng, Bingyu Li, Mingxun Zhou},
    hidelinks
}

\begin{document}

\maketitle
\suppressfloats[t]
\vspace{-10pt}

\begingroup
\setlength{\parskip}{4pt plus 1pt minus 1pt}
\begin{abstract}
    Relay and reseller APIs mediate access to large language models (LLMs), but users cannot directly verify which model serves them. We introduce \name, a black-box auditing protocol based on stable factual recall near the knowledge boundary, including repeatable wrong answers. KBF generates benign, renewable probes and calibrates audit decisions against reference self-variation. Across 16 production endpoints, KBF detects all 155 economically relevant substitutions without rejecting any of the 16 same-reference controls. KBF remains robust to deployment variation and reaches 95\% TPR at a substitution rate as low as 15\% in mixed-routing simulations. Field audits flag 7 of 28 endpoints across six platforms as statistically inconsistent with their references. After reference enrollment, even GPT-6 Astra costs only approximately \$0.67 per online audit at the recorded API prices.

\end{abstract}
\vspace{-4pt}
\setlength{\parskip}{2pt plus 0.5pt minus 0.5pt}
\setlength{\textfloatsep}{8pt plus 1pt minus 1pt}
\setlength{\floatsep}{6pt plus 1pt minus 1pt}
\captionsetup{skip=3pt}

\vspace{-2pt}
\section{Introduction}
\vspace{-2pt}
Relay and reseller APIs commonly provide access to large language models (LLMs). Users pay a third-party relay for an advertised model, while the relay controls which upstream endpoint serves each request. Users cannot directly verify which model serves them. The price gap between flagship and budget models creates an incentive for silent substitution. A relay can reduce its serving costs by routing some or all requests to a cheaper backend while continuing to charge for the advertised model. Measurement studies have reported deceptive model claims in shadow APIs~\cite{zhang2026real} and malicious intermediaries in the LLM supply chain~\cite{liu2026your}.

This motivates a technical question. Given only black-box access to an official reference API and a suspect relay API, can an auditor detect model substitution? The auditor can send prompts to these endpoints and read their responses, but methods requiring model internals or provider cooperation fall outside this setting~\cite{zhu2026auditing,yang2024fingerprint,sun2024zkllm}. The audit must remain robust to legitimate deployment variations, including changes to system prompts, decoding settings, and serving backends, among others.

\begin{table}[tp]
    \centering
    \footnotesize
    \captionsetup{skip=4pt}
    \setlength{\tabcolsep}{1.1pt}
    \caption{Black-box auditors compared on accuracy (Acc.), robustness to deployment variation (Rob.), benign and renewable probes (Ben.), and economic practicality (Econ.). --- denotes an untested setting.}
    \label{tab:requirements}
    \begin{tabular}{@{}lcccc@{}}
        \toprule
        \textbf{Method} & \textbf{Acc.} & \textbf{Rob.} & \textbf{Ben.} & \textbf{Econ.} \\
        \midrule
        \textbf{\name (Ours)} & \ding{51} & \ding{51} & \ding{51} & \ding{51} \\
        LLMmap~\cite{llmmap} & \ding{55}     & \ding{55} & \ding{55}     & \ding{51} \\
        MET~\cite{met} & \ding{51} & \ding{55} & \ding{51} & \ding{55}     \\
        ZeroPrint~\cite{zeroprint} & \ding{55}     & \ding{55} & \ding{51} & \ding{55}     \\
        IRIS~\cite{zhang2026model} & \ding{51} & \ding{55} & \ding{55}     & \ding{51} \\
        GateScope~\cite{lin2026behavioral} & \ding{51}     & --- & \ding{55}     & \ding{55}     \\
        \bottomrule
    \end{tabular}
\end{table}
A useful audit protocol should satisfy four requirements. (1) \textbf{Accuracy} requires detecting economically meaningful substitutions while keeping false-positive rates low, since false alarms can implicate honest relays. (2) \textbf{Robustness to deployment variation} requires reliable audit decisions despite benign differences in system prompts, temperature, output limits, retrieval wrappers, and serving backends. (3) \textbf{Benign and renewable probes} should resemble ordinary user requests and be inexpensive to regenerate, since a relay can filter recognizable probes or selectively forward them to the genuine backend. (4) \textbf{Economic practicality} requires low probe-generation and online audit costs to support repeated auditing.

Existing black-box auditors do not yet meet these requirements jointly in our setting (Table~\ref{tab:requirements}). Fixed probes can be recognized and handled selectively~\cite{llmmap,lin2026behavioral}. Output distributions and response-based fingerprints can also change with prompts, decoding, or serving implementations, complicating substitution detection~\cite{met,zeroprint,zhang2026model}. Repeated sampling and costly probes further limit frequent auditing~\cite{met,lin2026behavioral}. These limitations motivate an audit based on renewable, ordinary-looking probes whose reference behavior remains stable under benign deployment variation.

\Needspace{18\baselineskip}
\begin{wrapfigure}{l}{0.55\textwidth}
    \vspace{-2pt}
    \centering
    \captionsetup{font=small,skip=4pt}
    \includegraphics[width=\linewidth,trim=70bp 459bp 68bp 48bp,clip]{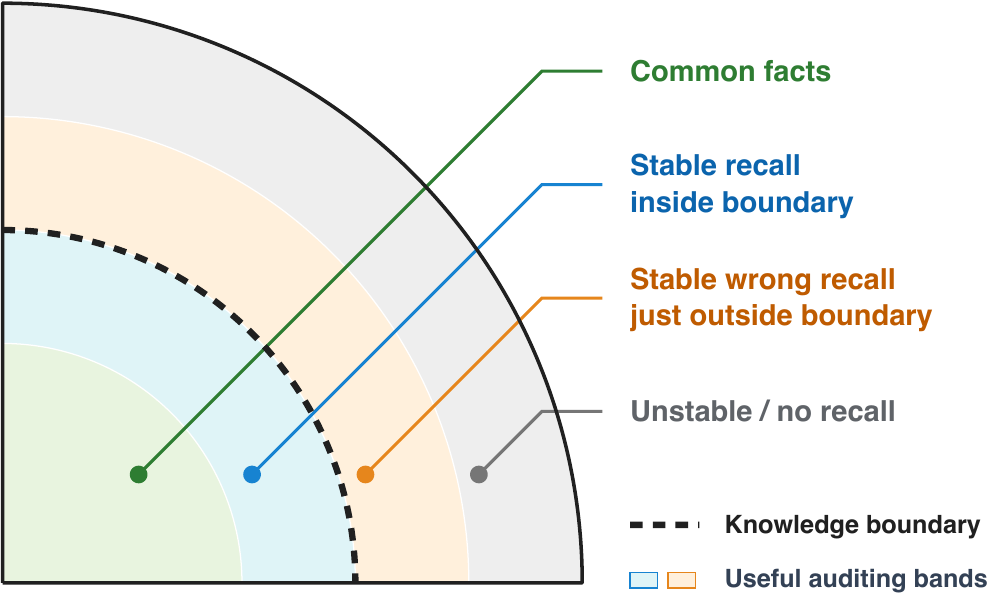}
    \caption{Knowledge-boundary recall as an audit signal. KBF retains stable reference answers, including repeatable wrong answers just outside the boundary.}
    \Description{Quarter-circle shell diagram showing common facts, stable recall inside the boundary, stable wrong recall just outside the boundary, and unstable no-recall facts far beyond the boundary.}
    \label{fig:kbf_boundary_intuition}
    \vspace{-20pt}
\end{wrapfigure}
\noindent Our insight is that \textit{factual recall near the knowledge boundary can provide both reference stability and model discrimination} (Figure~\ref{fig:kbf_boundary_intuition}).
Models often share common facts, while obscure queries can make the reference inconsistent. Between these extremes, the reference may repeat a specific answer that other endpoints do not reproduce. This includes stable factual errors, which remain useful because the audit tests agreement with
the reference rather than correctness.

We design \emph{Knowledge-Boundary Fingerprinting} (KBF), a black-box relay-auditing protocol based on this signal. KBF has three stages (Figure~\ref{fig:system_model}). (1) During offline probe generation, the auditor generates candidate probes through the official reference API and screens them for stable reference answers and, when useful, disagreement with a cheaper contrast model. (2) During reference self-calibration, the auditor re-queries retained probes and derives an upper confidence bound on reference self-disagreement. (3) During the online audit, KBF compares suspect responses with the stored reference answers under domain-specific matching rules and applies a one-sided binomial test against the calibrated bound. Probes can be refreshed per reference version.

\Needspace{18\baselineskip}
\begin{wrapfigure}{r}{0.56\textwidth}
    \vspace{-10pt}
    \centering
    \captionsetup{font=small,skip=4pt}
    \includegraphics[width=\linewidth]{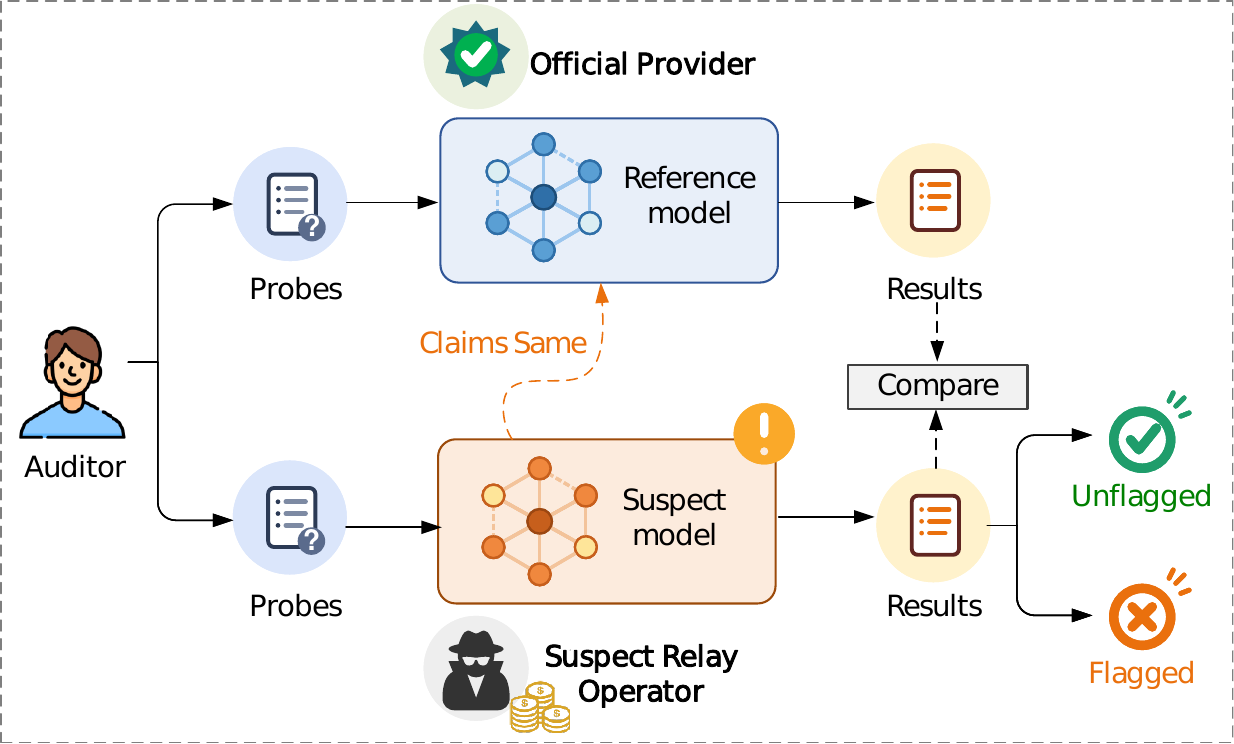}
    \caption{Black-box audit flow for KBF. The auditor compares a claimed reference endpoint with a suspect relay endpoint using only public API access.}
    \Description{Diagram of an auditor querying an official reference endpoint and a suspect relay endpoint, with the suspect relay controlling its backend serving logic.}
    \label{fig:system_model}
    \vspace{-8pt}
\end{wrapfigure}

KBF combines accurate substitution detection and robustness to deployment variation with benign, renewable probes and low audit costs. Across 16 production LLM endpoints from eight model families, KBF detects all 155 economically relevant substitutions without rejecting any of the 16 same-reference controls (Section~\ref{sec:detection}). Under six deployment configurations on five representative models, it detects all 60 substitutions with no false positives in 30 same-reference trials, whereas none of the four evaluated baselines achieves both outcomes (Section~\ref{sec:config-robustness}). In our latest evaluation, even GPT-6 Astra requires only approximately \$0.67 per audit at the recorded OpenRouter prices (Section~\ref{sec:audit-cost}). In mixed-routing simulations, our routing method reaches 95\% TPR at a substitution rate as low as 15\% (Section~\ref{sec:adaptive-routing}). In the six-platform field-study subset, KBF flags 7 of 28 endpoints as inconsistent with their references, predominantly for premium proprietary models, supporting further investigation of real relay APIs (Section~\ref{sec:shadow-api}).

\vspace{-4pt}
Our contributions are as follows.
\par\vspace{-4pt}
\begin{enumerate}[leftmargin=1.5em,topsep=0pt,itemsep=2pt,parsep=0pt,partopsep=0pt]
    \item \textbf{Knowledge-boundary auditing.} We develop KBF, a black-box protocol based on stable factual recall near the knowledge boundary, including repeatable wrong answers. KBF combines accurate, robust detection with benign, renewable probes and low audit costs for practical auditing of real-world APIs.
    \item \textbf{Systematic evaluation.} We compare KBF with existing auditors across production models and deployment settings, characterize the stability and discrimination of boundary recall, and use ablations to examine the contributions of probe generation and screening.
    \item \textbf{Mixed routing and real-world audits.} We demonstrate partial-substitution detection and routing-fraction estimation under the evaluated routing model. Real relay audits reveal endpoints statistically inconsistent with their references, predominantly for premium proprietary models.
    \item \textbf{Practical auditing tools.} We provide runnable scripts for probe generation and API auditing at \url{https://github.com/Ooo0ption/KBF}.
\end{enumerate}

\vspace{-5pt}
\section{Related Work}
\vspace{-2pt}
\label{sec:related}

\partitle{Auditing black-box LLM APIs}
Existing API auditors use behavioral signatures and statistical tests to check model claims. LLMmap and GateScope identify enrolled models from responses to fixed probe sets using learned representations or classifiers~\cite{llmmap,lin2026behavioral}. MET compares completion distributions from two endpoints with a two-sample test~\cite{met}. RUT tests whether API responses are consistent with a reference distribution through rank-based scores, requiring local access to the reference model~\cite{zhu2026auditing}. ZeroPrint measures how response embeddings change under input perturbations and compares these local-sensitivity fingerprints~\cite{zeroprint}. IRIS uses random-generation bias to detect model substitution and partial routing, estimate the routed fraction, and select an audit budget~\cite{zhang2026model}. For relay auditing, the comparison must account for benign deployment variation between the official reference and an honest relay. \name uses stable factual recall near the knowledge boundary of a claimed reference endpoint, with a binomial decision rule calibrated on reference self-variation. Appendix~\ref{app:baselines-related} covers baseline configurations, detailed comparisons, and approaches requiring stronger access or provider cooperation.

\partitle{Knowledge boundaries}
Knowledge-boundary research studies which facts a model can recall reliably and how it should respond when reliable recall fails~\cite{li2025knowledge,deng2025unveiling,zhou2026retrieval}. Concurrent work on Incompressible Knowledge Probes (IKP) uses factual recall and shared incorrect answers to estimate model capacity and examine model relationships~\cite{li2026incompressible}. IKP uses a common question bank to characterize models. \name generates probes for a claimed reference and uses them to audit endpoint consistency. Its probes can be regenerated privately and are screened for stable responses across reference configurations. Both approaches independently exploit model-specific factual recall, including repeatable errors, as an observable signal. Our audit tests whether a suspect reproduces this behavior under the evaluated deployment variations.

\vspace{-5pt}
\section{Problem Setting}
\vspace{-2pt}
\begingroup
\setlength{\parskip}{2pt plus 0.5pt minus 0.5pt}
\setlength{\abovedisplayskip}{3pt plus 1pt minus 1pt}
\setlength{\belowdisplayskip}{3pt plus 1pt minus 1pt}
\setlength{\abovedisplayshortskip}{0pt plus 1pt}
\setlength{\belowdisplayshortskip}{3pt plus 1pt minus 1pt}
\label{sec:prelim}

We study whether a third-party relay API serves the model it advertises.

\partitle{Endpoints}
The auditor has black-box access to two randomized oracles, the official reference $\mathcal{O}_{\mathrm{ref}}(x,c)$ and suspect relay $\mathcal{O}_{\mathrm{sus}}(x,c)$, where $x$ is a user query and $c\in\mathcal{C}$ is a requested deployment configuration (Figure~\ref{fig:system_model}). We assume the reference applies these requests up to ordinary provider-side nondeterminism. The auditor requests the same configurations from the suspect but cannot assume compliance. The auditor has no access to weights, logits, training data, inference code, provider metadata, or routing logs. KBF fingerprints the served API endpoint, not isolated parametric weights.

\partitle{Auditing goal}
A protocol $\mathsf{Audit}$ adaptively queries both endpoints and outputs \textsc{Same} or \textsc{Different}. The hypotheses are
\[
    H_0: \mathcal{O}_{\mathrm{sus}} \approx \mathcal{O}_{\mathrm{ref}}
    \quad\text{vs.}\quad
    H_1: \mathcal{O}_{\mathrm{sus}} \not\approx \mathcal{O}_{\mathrm{ref}}.
\]
Here $\approx$ is operational: the suspect is consistent with the reference only if its audited behavior remains within the measured reference self-variation (including reasonable variation due to wrappers, temperature, and backend deployment). Appendix~\ref{app:problem-setting} describes the audit scope and non-goals.

\partitle{Relay capabilities}
The adversary is the suspect relay operator. It controls backend selection, decoding parameters, hidden prompts, wrappers, and quantization. Mixed routing gives
\[
    \mathcal{O}_{\mathrm{sus}} =
    \begin{cases}
        \mathcal{O}_{\mathrm{ref}} & \text{with probability } 1-\pi, \\
        \mathcal{O}_{\mathrm{sub}} & \text{with probability } \pi.
    \end{cases}
\]
Here $\mathcal{O}_{\mathrm{sub}}$ is a substitute backend and $\pi\in[0,1]$ is the substitution rate, with $\pi=1$ giving full substitution. Routing may depend on the query, requested configuration, load, user account, or request timing. The adversary may know the algorithm and public probe sets and query the official reference endpoint. It cannot control that endpoint, the auditor's local computation, or auditor--API communication. KBF does not provide cryptographic security against perfect probe recognition.

\partitle{Evaluation criteria}
We measure the false-positive rate (FPR) and true-positive rate (TPR),
\[
    \begin{aligned}
    \mathrm{FPR}
        &= \Pr[\mathsf{Audit}(\mathcal{O}_{\mathrm{ref}},\mathcal{O}_{\mathrm{sus}})=\textsc{Different} \mid H_0],\\
    \mathrm{TPR}
        &= \Pr[\mathsf{Audit}(\mathcal{O}_{\mathrm{ref}},\mathcal{O}_{\mathrm{sus}})=\textsc{Different} \mid H_1].
    \end{aligned}
\]
FPR is the primary safety metric because a false accusation against an honest provider is costly. For mixed routing, TPR depends on the substitution rate $\pi$ and the substitute model. Robustness requires low reference self-noise across repeated queries and configurations.
\par\endgroup

\begingroup
\setlength{\parskip}{2pt plus 0.5pt minus 0.5pt}
\setlength{\abovedisplayskip}{3pt plus 1pt minus 1pt}
\setlength{\belowdisplayskip}{3pt plus 1pt minus 1pt}
\setlength{\abovedisplayshortskip}{0pt plus 1pt}
\setlength{\belowdisplayshortskip}{3pt plus 1pt minus 1pt}
\setlength{\intextsep}{8pt plus 2pt minus 2pt}
\vspace{-5pt}
\section{Method}
\vspace{-2pt}
\label{sec:construction}

\vspace{-3pt}
\subsection{Boundary Recall as an Audit Signal}
\vspace{-2pt}
\label{sec:boundary-signal}

Factual recall near the knowledge boundary can remain stable for the reference while differing across models.
Interior probes concern common facts shared across models.
Beyond-boundary probes are too obscure for the reference to answer consistently.
Between these extremes, boundary probes elicit answers that the reference repeats reliably but other models reproduce less often.

KBF generates and screens boundary probes as described in Section~\ref{sec:kbf}.
To examine this signal, we collect all three probe types for each of four reference models and audit every model on each pool.
Figure~\ref{fig:e16-boundary-signal} compares their agreement with recorded reference answers, showing how boundary probes combine reference stability with model discrimination. Stable factual errors can also contribute because the audit tests reference agreement rather than correctness.

\begin{figure}[!htbp]
    \centering
    \includegraphics[width=0.88\linewidth]{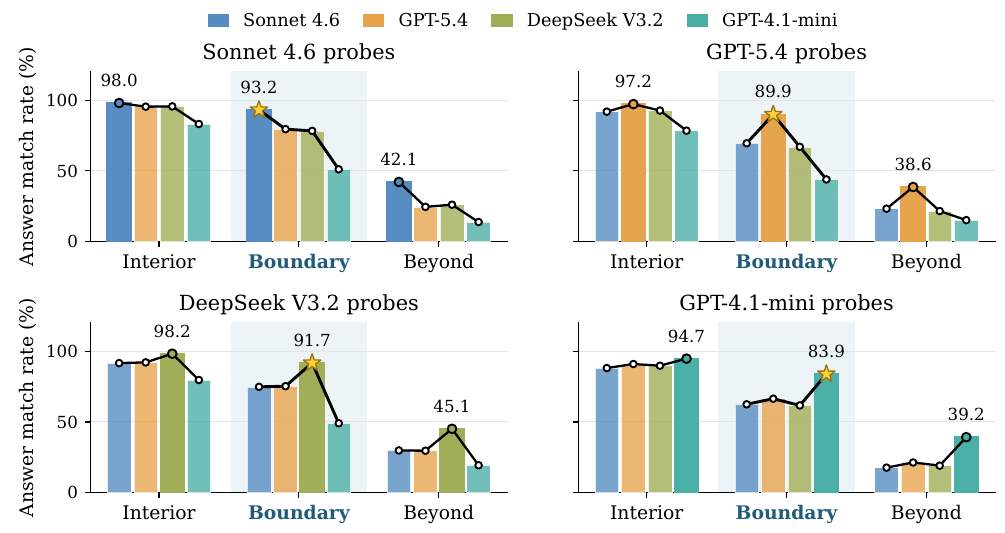}
    \caption{Match rates by reference probe set. Dark bars and labels show reference self-evaluation, lines connect models within groups, and stars mark boundary peaks. Rates compare separate post-construction responses with recorded reference answers, excluding unscored responses.}
    \label{fig:e16-boundary-signal}
    \vspace{-8pt}
\end{figure}

\partitle{Reference stability}
Boundary probes retain 83.86\%--93.15\% reference agreement on fresh queries, providing a reproducible audit baseline.
Interior recall is stable, but other models match these answers at 78.43\%--95.58\%, limiting model discrimination.
Repeatability alone is insufficient for auditing.

\partitle{Model discrimination}
Other models reproduce boundary answers at rates of 43.91\%--79.48\%, below the corresponding reference in all 12 directed model pairs.
Beyond-boundary probes also show cross-model differences, but reference agreement drops to 38.57\%--45.11\%, making those differences harder to separate from reference inconsistency.

\Needspace{7\baselineskip}
\vspace{-5pt}
\subsection{Knowledge Boundary Fingerprinting}
\vspace{-2pt}
\label{sec:kbf}

\partitle{Stage I. Reference probe generation}
\label{phase1}
Our numerical-probe implementation generates candidates independently within domains.
Each domain $d$ specifies a prompt template, a valid numerical range, and a comparison rule $\mathsf{match}_d$.
The same $\mathsf{match}_d$ applies to reference-consistency checks, self-calibration, and auditing.
Unparseable or out-of-range values are invalid.

For each domain, KBF generates candidates over rounds indexed by $t$.
Each round asks the reference endpoint to propose structured candidate records, typically as \texttt{name | value} pairs.
The prompt specifies a domain theme and a short exclusion list of previously proposed names.
KBF parses the proposed records, removes duplicate names, rejects values outside the domain range, and renders the remaining records into short audit prompts $q_i$.

Reference-consistency checks $\mathcal{C}_{\mathrm{cons}}$ re-query each candidate under benign configuration changes, such as prompt variants or decoding settings.
For each prompt $q_i$ in domain $d_i$, KBF parses the responses into domain-valid values, denoted by $A_i$.
The candidate survives only when all responses are valid and agree under $\mathsf{match}_{d_i}$, and KBF records their consensus value as $a_i$.
The fingerprint stores the $N$ surviving probes as
\[
    P_{\mathrm{ref}}=\{(q_i,d_i,a_i,\mathsf{match}_{d_i})\}_{i=1}^{N}.
\]
KBF tracks the number of surviving probes per round and stops searching a domain when further rounds yield no stable probes.

\partitle{Stage II. Reference self-calibration}
\label{phase2}
KBF fixes $P_{\mathrm{ref}}$ and re-queries the reference endpoint under the audit configuration $c_{\mathrm{audit}}$, yielding parsed values $\tilde{a}_i$.
Missing, unparseable, and out-of-range answers are excluded from scoring.
For any domain-valid returned value $v$, define the discrepancy indicator
\[
    \delta_i(v)=
    \begin{cases}
        0, & \text{if }\mathsf{match}_{d_i}(a_i,v)=1,\\
        1, & \text{otherwise}.
    \end{cases}
\]
Let $n_{\mathrm{self}}$ count valid reference responses and $k_{\mathrm{self}}$ their discrepancies.
KBF calibrates the null bound as
\[
    p_0=\mathrm{CP}_{99}(k_{\mathrm{self}},n_{\mathrm{self}}).
\]
Here $\mathrm{CP}_{99}$ is the one-sided 99\% Clopper--Pearson upper confidence bound on the reference discrepancy rate~\cite{clopper1934use}.
We use this confidence level, $\gamma=0.99$, unless stated otherwise.
The fingerprint passed to online auditing is $F_{\mathrm{ref}}=(P_{\mathrm{ref}},k_{\mathrm{self}},p_0)$.
The binomial model treats scored probes as independent discrepancy trials with a common reference error rate.

\partitle{Stage III. Suspect endpoint audit}
KBF sends each enrolled prompt to the suspect endpoint $\mathcal{O}_{\mathrm{sus}}$ under $c_{\mathrm{audit}}$ and parses the returned value $\hat{a}_i$.
For each valid response, KBF records $Z_i=\delta_i(\hat{a}_i)$ and counts discrepancies as $k$.
The audit denominator $n$ counts valid responses, excluding missing, unparseable, and out-of-range answers.
Under the binomial model, let $p_{\mathrm{sus}}$ denote the suspect discrepancy probability on the scored probes.
KBF tests whether it exceeds the calibrated reference bound,
\[
    H_0:p_{\mathrm{sus}}\leq p_0,
    \qquad H_1:p_{\mathrm{sus}}>p_0.
\]
KBF computes the one-sided probability of at least $k$ discrepancies under the null bound,
\[
    p_{\mathrm{audit}}=\Pr(X\geq k),\qquad X\sim\mathrm{Binomial}(n,p_0).
\]
If $p_{\mathrm{audit}}<\alpha=0.05$, KBF rejects $H_0$ and reports \textsc{Different}.
Otherwise, it reports \textsc{Same}, meaning the audit did not find enough discrepancies to exceed the reference-noise bound.

Appendix~\ref{sec:kbf-implementation} provides the full procedures and implementation details for all three stages and discusses the implications of probe dependence.
In our ablations, specialist generation yields more retained probes, stability filtering reduces reference noise, and contrast filtering improves model discrimination (Appendix~\ref{app:signal-ablation}).

\vspace{-5pt}
\subsection{Partial-Routing Extension}
\vspace{-2pt}
\label{sec:kbf-routing-overview}
\label{sec:kbf-adaptive-routing}

Each API request contains up to ten probes from one domain.
The relay independently routes the entire request to substitute $S$ with probability $\pi$, or to the reference otherwise.
KBF fixes the reference-matched scoring set $\mathcal{I}_0=\{i\mid\tilde{a}_i\text{ is valid and }\delta_i(\tilde{a}_i)=0\}$ and preserves the original request groups.
Let $B$ count groups with scored probes, $A_b$ be the mismatch fraction from a subsequent reference check, and $Y_{b1},Y_{b2}$ the fractions from two online rounds.
Each batch contributes one observation, allowing arbitrary dependence among its scored probes.

\partitle{Paired detection}
We use two audit rounds to strengthen the detection signal, with one-round results reported in Appendix~\ref{app:adaptive-routing-details}.
KBF queries each batch twice and compares both responses with the stored reference check using
\[
    T=\sum_{b=1}^{B}\left[
       \operatorname{sgn}(Y_{b1}-A_b)
       +\operatorname{sgn}(Y_{b2}-A_b)\right].
\]
It computes the null distribution of $T$ by uniformly reassigning the reference label among the three records within each batch.
Both online rounds share one reference record and yield one joint decision.
KBF rejects when the upper-tail probability is below $0.05$.
Validity requires independent batch groups and exchangeable reference and honest-relay records within each group.
Adjacent reference checks assess temporal variation.

\partitle{Routing-fraction estimate}
Let $q_{Sb}$ be the mismatch fraction of a fixed substitute fingerprint on batch $b$.
Write $\bar Y_b=(Y_{b1}+Y_{b2})/2$, $d_b=\bar Y_b-A_b$, and $g_b=q_{Sb}-A_b$, with bars denoting averages over batches.
Here $d_b$ measures the observed increase over the reference, and $g_b$ the increase under full substitution.
Under stable reference means, $\bar d\approx\pi\bar g$, giving the point estimate
\[
    \widehat\pi_S=\operatorname{clip}_{[0,1]}\!\left(\frac{\bar d}{\bar g}\right),
    \qquad \bar g>0.
\]
Clipping restricts the estimate to $[0,1]$, and small mismatch gaps make the estimate sensitive to sampling variation.

\partitle{Unknown substitute}
For each candidate $S\in\mathcal C$, KBF recomputes $\bar g$ from its fingerprint and applies the same estimator to obtain $\widehat\pi_S$.
When all candidate mean mismatch gaps are positive, it reports
\[
    \widehat\Pi_{\mathcal C}=\left[
        \min_{S\in\mathcal C}\widehat\pi_S,\;
        \max_{S\in\mathcal C}\widehat\pi_S
    \right].
\]
This range reflects uncertainty about substitute identity and is not a confidence interval.
Its interpretation requires the true substitute to belong to the pool.
Content-dependent or audit-aware routing can prevent these estimates from representing the traffic-wide substitution fraction.

\par\endgroup

\begingroup
\setlength{\parskip}{2pt plus 0.5pt minus 0.5pt}
\setlength{\textfloatsep}{8pt plus 1pt minus 1pt}
\setlength{\floatsep}{6pt plus 1pt minus 1pt}
\setlength{\intextsep}{8pt plus 2pt minus 2pt}
\captionsetup{skip=3pt}
\vspace{-5pt}
\section{Experiments}
\vspace{-2pt}
\label{sec:evaluation}

We evaluate substitution detection, enrollment and audit cost, robustness to deployment and probe-format changes, partial routing, and real-world relay endpoints.

\vspace{-5pt}
\subsection{Experimental Setup}
\vspace{-2pt}
\label{sec:setup}

We evaluate KBF in controlled settings where the claimed endpoint, provider route, probe set, and decision rule are fixed.
This design isolates model behavior from avoidable routing noise and provides a baseline for the field audit.

\partitle{Models}
Table~\ref{tab:models} summarizes the endpoint pool.
We evaluate 16 production LLM API endpoints across eight model families and three price tiers, comprising six T1 flagship, five T2 mid-range, and five T3 budget endpoints.
OpenRouter~\cite{openrouter} input prices recorded in March 2026 range from \$0.05 to \$5.00 per million tokens.

\partitle{Reference endpoints}
For each claimed model, we enroll the reference fingerprint through OpenRouter with the provider fixed to the entry in Table~\ref{tab:models} and fallback routing disabled.

\partitle{Baselines}
We compare KBF with five black-box auditors using their released implementations.
In Section~\ref{sec:config-robustness}, KBF and the evaluated baselines use shared deployment configurations.
The baselines are LLMmap~\cite{llmmap}, MET~\cite{met}, ZeroPrint~\cite{zeroprint}, IRIS~\cite{zhang2026model}, and GateScope~\cite{lin2026behavioral}.
Appendix~\ref{app:baseline-details} describes the implementations, audit adaptations, and evaluation coverage.

\partitle{Metrics}
We count false positives (FP, lower is better), substitution alarms in same-reference trials, and true positives (TP, higher is better), correctly detected substitutions.

\begin{table}[!htbp]
    \centering
    \footnotesize
    \setlength{\tabcolsep}{2.5pt}
    \caption{The 16 evaluated LLM endpoints. \emph{Provider} identifies the pinned OpenRouter route. \emph{\#Probes} counts retained numerical probes. \emph{Self-Err} is the reference calibration self-disagreement rate, whose $\mathrm{CP}_{99}$ upper bound defines $p_0$. The diagonal of Figure~\ref{fig:heatmap_detection} reports independent same-reference retests. $^{*}$Input/output prices are in USD per million tokens, recorded from OpenRouter in March 2026.}
    \label{tab:models}
    \begin{tabular}{llll | rr | cc}
        \noalign{\hrule height 1.5pt}
        \multirow{2}{*}{\textbf{Tier}} & \multirow{2}{*}{\textbf{Model}} & \multirow{2}{*}{\textbf{Model Family}} & \multirow{2}{*}{\textbf{Model Provider}} &
        \multicolumn{2}{c|}{\textbf{Price$^{*}$ (\$/M Tokens)}} &
        \multirow{2}{*}{\textbf{\#Probes}} & \multirow{2}{*}{\textbf{Self-Err}} \\
        \cline{5-6}
        & & & & \textbf{Input} & \textbf{Output} & & \\
        \hline
        T1 & Claude Opus 4.6   & Anthropic & Amazon Bedrock &  5.00 & 25.00 & 681 & 4.3\% \\
        T1 & Claude Sonnet 4.6 & Anthropic & Google         &  3.00 & 15.00 & 224 & 1.3\% \\
        T1 & GPT-5.4           & OpenAI    & OpenAI         &  2.50 & 10.00 & 317 & 1.6\% \\
        T1 & Gemini 3 Flash    & Google    & Google         &  0.50 &  2.50 & 315 & 2.2\% \\
        T1 & GLM-5             & Z.AI      & Z.AI           &  0.72 &  2.20 & 405 & 4.2\% \\
        T1 & Qwen3.5-397B-A17B & Alibaba   & Alibaba        &  0.39 &  1.20 & 243 & 1.7\% \\
        \hline
        T2 & DeepSeek-V3.2  & DeepSeek & Google      & 0.26 & 0.42 & 364 & 3.3\% \\
        T2 & GPT-4.1-mini   & OpenAI   & OpenAI      & 0.40 & 1.60 & 134 & 6.0\% \\
        T2 & GLM-4.7        & Z.AI     & Z.AI        & 0.38 & 2.00 & 346 & 4.6\% \\
        T2 & Kimi-K2-0905   & Moonshot & Moonshot AI & 0.40 & 2.50 & 300 & 4.7\% \\
        T2 & Qwen3.5-27B    & Alibaba  & Alibaba     & 0.20 & 0.30 & 115 & 4.3\% \\
        \hline
        T3 & GPT-4.1-nano          & OpenAI  & OpenAI    & 0.10 & 0.40 & 109 &  7.3\% \\
        T3 & LLaMA-4-Scout         & Meta    & Groq      & 0.08 & 0.30 & 146 & 11.7\% \\
        T3 & Qwen3.5-9B            & Alibaba & Together  & 0.05 & 0.10 & 105 &  3.8\% \\
        T3 & GLM-4.7-Flash         & Z.AI    & DeepInfra & 0.06 & 0.20 & 309 & 16.2\% \\
        T3 & Gemini 2.5 Flash Lite & Google  & Google    & 0.10 & 0.40 & 210 & 13.8\% \\
        \noalign{\hrule height 1.5pt}
    \end{tabular}
\end{table}

\vspace{-5pt}
\subsection{Detection Accuracy}
\vspace{-2pt}
\label{sec:detection}
\label{sec:audit-cost}

We audit all pairs in the 16-model pool.
KBF flags all \textbf{155 economically relevant substitutions} at $p<0.05$ under the $\mathrm{CP}_{99}$ binomial test, while producing no false-positive rejection in the 16 same-reference trials.

\partitle{Economically Relevant Pairs}
A pair $(M_r,M_t)$ is economically relevant when target endpoint $M_t$ is cheaper than claimed reference $M_r$ or both are in the same price tier.
Lower-tier substitutions capture direct cost savings.
Same-tier substitutions capture smaller incentives from within-tier price differences, easier access, compliance burden, and regional availability.
We exclude only cross-tier upgrades to a more expensive target, such as T2$\rightarrow$T1 and T3$\rightarrow$T2/T1, from the threat model and TPR/FPR estimates.

\Needspace{6\baselineskip}
\partitle{Substitution Detection}
Figure~\ref{fig:heatmap_detection} shows the detection matrix, with white off-diagonal cells marking economically irrelevant upgrade directions.
KBF detects all \textbf{12/12} within-family economically relevant substitutions and all \textbf{143/143} cross-family substitutions, showing that the signal extends beyond visibly different model families.
KBF also detects most upgrade substitutions.
In September 2026, we generated fresh probe sets for an expanded model pool that includes newly released models (Appendix~\ref{app:detection-extended}).

\begin{figure}[!htbp]
    \vspace{-4pt}
    \centering
    \includegraphics[width=0.85\textwidth]{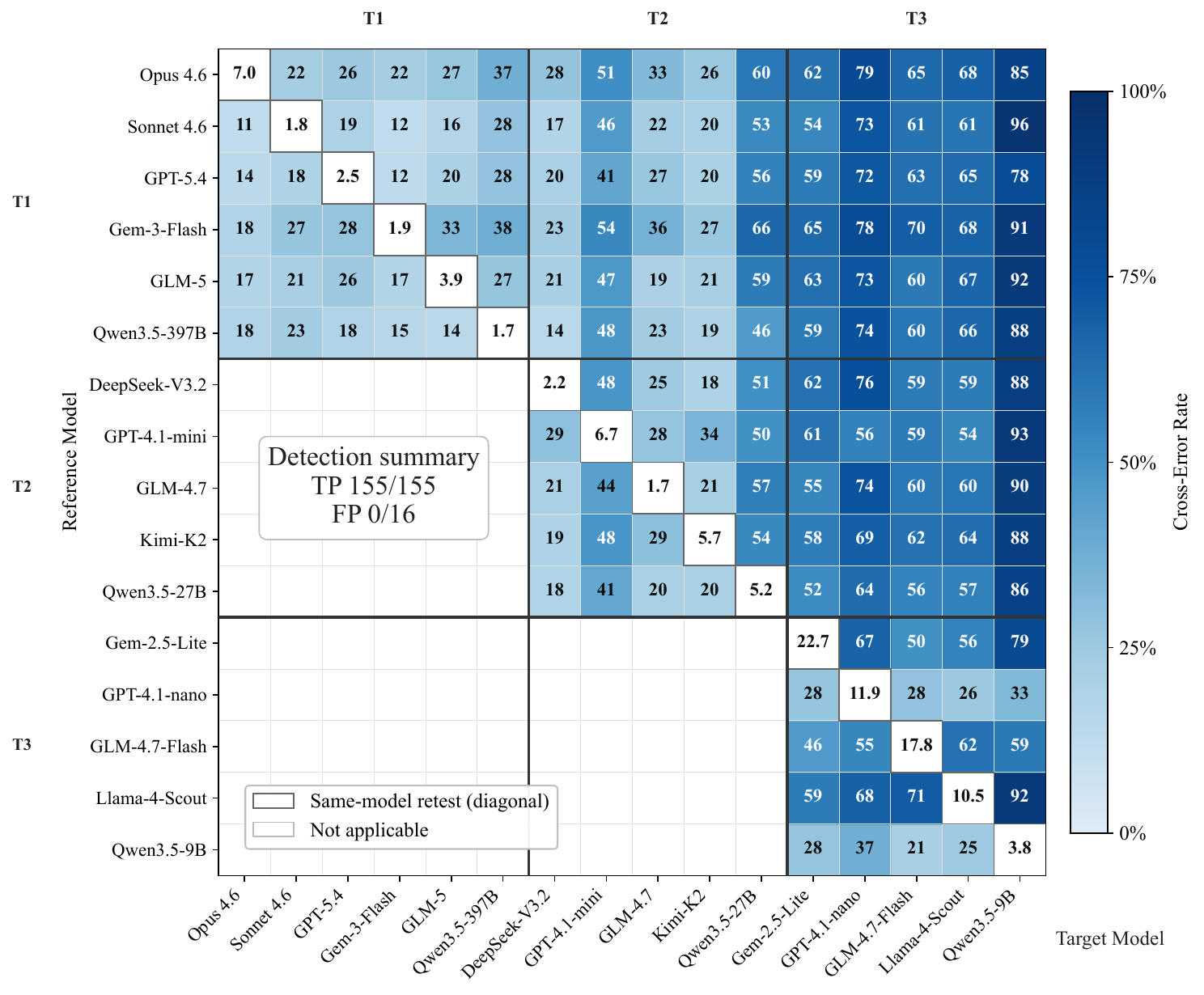}
    \caption{Detection heatmap over 16$\times$16 model pairs. Color indicates the mismatch rate on the reference probe set. Diagonal cells show independent same-reference retests evaluated against their calibration bounds. White off-diagonal cells denote economically irrelevant upgrade directions.}
    \Description{A 16 by 16 heatmap of model-pair mismatch rates. Diagonal cells are same-reference trials. All economically relevant substitution directions are highlighted and detected, while upgrade directions are left blank.}
    \label{fig:heatmap_detection}
    \vspace{-6pt}
\end{figure}

\Needspace{145pt}
\begin{wraptable}{r}{0.5\textwidth}
    \vspace{0pt}
    \small
    \setlength{\tabcolsep}{5pt}
    \caption{Same-reference false positives.}
    \label{tab:fp_baseline}
    \begin{tabular}{lcc}
        \toprule
        \textbf{Method} & \textbf{FP} & \textbf{FPR} \\
        \midrule
        \textbf{KBF (ours)}        & \textbf{0 / 16} & \textbf{0.0\%} \\
        LLMmap~\cite{llmmap}       & 5 / 16          & 31.3\% \\
        MET~\cite{met}             & 1 / 16          & 6.3\% \\
        ZeroPrint~\cite{zeroprint} & 1 / 16          & 6.3\% \\
        IRIS~\cite{zhang2026model} & 4 / 16          & 25.0\% \\
        GateScope$^{*}$~\cite{lin2026behavioral} & 0 / 12      & 0.0\% \\
        \bottomrule
    \end{tabular}
    \par\smallskip
    \parbox{\linewidth}{\footnotesize\raggedright $^{*}$GateScope was evaluated on a 12-model subset.}
    \vspace{-10pt}
\end{wraptable}
\partitle{False-Positive Control}
\label{sec:fp-control}
KBF produces 0/16 false positives in independent same-reference trials, with baseline results in Table~\ref{tab:fp_baseline}.
New tests in September 2026 also yield 0/28 false positives on the expanded model pool with freshly enrolled probes (Appendix~\ref{app:frontier-sets}).
Section~\ref{sec:config-robustness} and Appendix~\ref{sec:robustness} report false-positive control results for KBF under deployment variation and temporal retesting.

\partitle{Detection Cost}
(1) \textbf{Probe generation.} A separate May 2026 run enrolled 3,696 probes for six T1 references for \$7.07 (Appendix~\ref{app:enrollment-cost}).
Probes can be reused across audits of the same reference version.
(2) \textbf{Online audit.} At March 2026 API prices, Claude Opus~4.6 has the highest single-audit cost among the original 16 models, at \$0.24.
Among the 13 models added in September 2026, GPT-6 Astra is the most expensive at \$0.67 per audit at the recorded prices.
Both figures exclude enrollment.

\vspace{-5pt}
\subsection{Robustness to Deployment Variation}
\vspace{-2pt}
\label{sec:config-robustness}

Relay APIs often serve application endpoints with role prompts, nonzero temperatures, retrieval context, or response wrappers for product workflows.
An audit should preserve accuracy under these wrappers, maintaining 0\% FPR in same-reference trials and 100\% TPR in substitution trials.

\partitle{Cross-Method Stress Test}
We compare four baselines on Claude Sonnet~4.6 (T1), GPT-5.4 (T1), GLM-4.7 (T2), GPT-4.1-mini (T2), and Qwen3.5-9B (T3), yielding 5 same-reference setups and 10 economically relevant substitution pairs.
KBF and all baselines use six deployment variants with shared role prompts, temperatures, and RAG settings.
Each method undergoes 30 same-reference and 60 substitution trials.

Table~\ref{tab:config_robustness_methods} shows that \textbf{KBF separates honest endpoints from substitutions across all six tested deployment variants}, while the baselines raise false alarms and, in some cases, miss substitutions.
The reduced accuracy may reflect the sensitivity of baseline detection methods to temperature and prompt-wrapping changes relative to their reference configurations.
Additional tests cover adversarial prompts, identity spoofing, quantization, provider variation, temporal drift, threshold choice, agent interfaces, and deployment variation across the full model pool.
KBF remains stable in most tested settings (Appendix~\ref{sec:robustness}).
\begin{table*}[h]
    \centering
    \footnotesize
    \renewcommand{\arraystretch}{1.08}
    \setlength{\tabcolsep}{2.2pt}
    \caption{Configuration-robustness comparison on five representative models under six shared deployment wrappers. }
    \label{tab:config_robustness_methods}
    \begin{tabular}{c>{\raggedright\arraybackslash}p{2.4cm}cc | cc | cc | cc | cc | cc}
        \noalign{\hrule height 1.5pt}
        \multirow{2}{*}{\textbf{ID}} & \multirow{2}{*}{\textbf{Role}} & \multirow{2}{*}{\textbf{Temp}} & \multirow{2}{*}{\textbf{RAG}} &
        \multicolumn{2}{c|}{\textbf{KBF (ours)}} &
        \multicolumn{2}{c|}{\textbf{MET}} &
        \multicolumn{2}{c|}{\textbf{LLMmap}} &
        \multicolumn{2}{c|}{\textbf{ZeroPrint}} &
        \multicolumn{2}{c}{\textbf{IRIS}} \\
        \cline{5-14}
        & & & & FP & TP & FP & TP  & FP & TP & FP & TP & FP & TP \\
        \hline
        C0 & No role prompt & 0 & No & \textbf{0/5} & \textbf{10/10} & 4/5 & 10/10 & 3/5 & 6/10  & 4/5 & 7/10 & 0/5 & 8/10 \\
        C1 & Finance compliance & 0.2 & No & \textbf{0/5} & \textbf{10/10} & 5/5 & 10/10 & 3/5 & 10/10 & 5/5 & 7/10 & 1/5 & 8/10 \\
        C2 & Enterprise policy & 0 & Yes & \textbf{0/5} & \textbf{10/10} & 5/5 & 10/10 & 3/5 & 8/10  & 4/5 & 7/10 & 1/5 & 8/10 \\
        C3 & Customer support & 0.3 & No & \textbf{0/5} & \textbf{10/10} & 5/5 & 10/10 & 3/5 & 7/10  & 3/5 & 8/10 & 1/5 & 8/10 \\
        C4 & Medical triage & 0.1 & Yes & \textbf{0/5} & \textbf{10/10} & 5/5 & 10/10 & 4/5 & 7/10  & 3/5 & 8/10 & 1/5 & 8/10 \\
        C5 & Academic writing & 0.7 & No & \textbf{0/5} & \textbf{10/10} & 5/5 & 10/10 & 3/5 & 8/10  & 4/5 & 8/10 & 2/5 & 8/10 \\
        \hline
        \multicolumn{4}{l|}{\textbf{Total}} & \textbf{0/30 } & \textbf{60/60} & 29/30  & 60/60 & 19/30  & 46/60 & 23/30& 45/60 & 6/30 & 48/60 \\
        \noalign{\hrule height 1.5pt}
    \end{tabular}
\end{table*}

\vspace{-5pt}
\subsection{Probe Format Variation}
\vspace{-2pt}
\label{sec:probe-recognizability}
\label{sec:probe-format}

\begin{wraptable}{r}{0.52\textwidth}
    \vspace{-12pt}
    \centering
    \footnotesize
    \setlength{\tabcolsep}{2pt}
    \captionsetup{font=footnotesize}
    \caption{Probe features. $\bullet$ = retained, $\circ$ = removed. Number denotes numerical answers.}
    \label{tab:probe-format-constraints}
    \begin{tabular}{@{}lcccc@{}}
        \toprule
        Variant & Number & Cloze & Prompt & Batch \\
        \midrule
        Entity cloze       & $\circ$ & $\bullet$ & $\bullet$ & $\bullet$ \\
        Numerical QA       & $\bullet$ & $\circ$ & $\bullet$ & $\bullet$ \\
        Natural batch QA   & $\bullet$ & $\circ$ & $\circ$ & $\bullet$ \\
        Single-question QA & $\bullet$ & $\circ$ & $\circ$ & $\circ$ \\
        \bottomrule
    \end{tabular}
    \vspace{-10pt}
\end{wraptable}
KBF relies on the knowledge-boundary questions retained by screening rather than on their presentation format.
Reformulating KBF probes reduces dependence on a fixed request format.
We evaluate four variants that change the answer type or remove cloze blanks, fixed system prompts and answer-format rules, or batching (Table~\ref{tab:probe-format-constraints}).
All four detect every tested substitution without false positives, showing that detection remains effective across these changes in request format.
KBF can renew probe domains, questions, and queried numerical facts through fresh generation and screening.
Starting from questions retained by KBF screening, we use a low-cost LLM, GPT-4.1-mini, to rewrite questions, verify semantic equivalence, and extract answers from free-text replies.
This allows the screened probes to take many surface forms while preserving their factual content.
Appendix~\ref{app:probe-formats} gives the results and request formats.

\vspace{-5pt}
\subsection{Partial-Routing Detection and Estimation}
\vspace{-2pt}
\label{sec:adaptive-routing}

\begin{wrapfigure}{r}{0.52\textwidth}
    \vspace{-12pt}
    \centering
    \includegraphics[width=\linewidth]{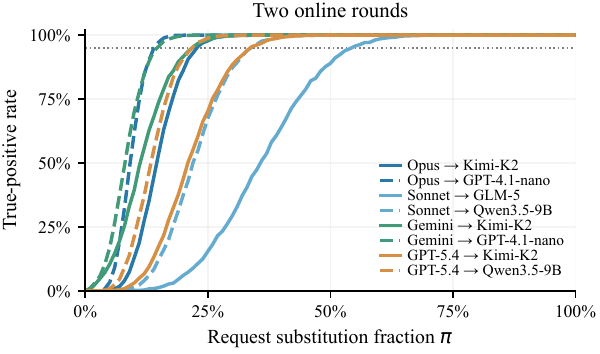}
    \captionsetup{font=small,skip=3pt}
    \caption{Two-round routing detection for eight pairs at $\alpha=0.05$. Solid and dashed curves denote same-class and budget substitutes. The dotted line marks 95\% TPR.}
    \label{fig:partial_routing_power}
    \vspace{-10pt}
\end{wrapfigure}
We evaluate eight reference--substitute pairs using the paired batch method in Section~\ref{sec:kbf-routing-overview}.
Each request contains up to ten probes and selects one backend for the entire batch, independently across requests and rounds.
We replay two-round audits using archived responses, measuring routing variability conditional on these records.

\partitle{Main Results}
Two rounds reach 95\% TPR at routing fractions of 22--55\% for same-class substitutes and 15--34\% for budget substitutes (Figure~\ref{fig:partial_routing_power}).
Additional experiments show that a second audit round improves detection and yields more precise routing-fraction estimates for known substitutes (Appendix~\ref{app:adaptive-routing-details}).

\vspace{-5pt}
\subsection{Real-World Relay Audits}
\vspace{-2pt}
\label{sec:shadow-api}

We audit eleven relay API platforms, anonymized as Platform~1--11, for statistical inconsistency with the reference probe-response distributions.
Table~\ref{tab:shadow_results} covers 28 available endpoints across six platforms.
Across all eleven platforms, flagged endpoints remain concentrated among Claude models (Appendix~\ref{app:shadow-details}).

\begingroup
\setlength{\intextsep}{6pt plus 1pt minus 1pt}
\begin{table}[!htbp]
    \centering
    \footnotesize
    \setlength{\tabcolsep}{2.5pt}
    \captionsetup{skip=4pt}
    \caption{Field audit of six anonymized relay platforms. Entries report mismatch rates, with \textcolor{green!50!black}{$\checkmark$} for no rejection, \textcolor{red!70!black}{$\times$} for rejection ($p<0.05$), and \textbf{?} near the decision threshold. Parentheses give reference self-error. Dashes denote unavailable endpoints. The Opus entry for Platform~5 uses the default tier.}
    \label{tab:shadow_results}
    \begin{tabular}{lccccccc}
        \toprule
        \textbf{Model (self-err)} & \textbf{P1} & \textbf{P2} & \textbf{P3} & \textbf{P4} & \textbf{P5} & \textbf{P11} & \textbf{Flagged/tested} \\
        \midrule
        Claude Opus 4.6 (4.3\%)        & \textcolor{red!70!black}{$\times$} 19.7\% & \textcolor{black}{\textbf{?}} 8.2\%  & ---                                       & \textcolor{green!50!black}{$\checkmark$} 4.3\%  & \textcolor{red!70!black}{$\times$} 47.9\% & \textcolor{black}{\textbf{?}} 9.0\%                                    & \textbf{4 / 5} \\
        Claude Sonnet 4.6 (1.3\%)      & \textcolor{red!70!black}{$\times$} 13.4\% & \textcolor{green!50!black}{$\checkmark$} 2.8\%  & \textcolor{green!50!black}{$\checkmark$} 2.7\%  & \textcolor{green!50!black}{$\checkmark$} 2.2\%  & \textcolor{red!70!black}{$\times$} 43.8\% & \textcolor{green!50!black}{$\checkmark$} 5.4\% & \textbf{2 / 6} \\
        GLM-4.7 (4.6\%)                & \textcolor{green!50!black}{$\checkmark$} 6.8\%  & \textcolor{red!70!black}{$\times$} 17.4\% & ---                                       & \textcolor{green!50!black}{$\checkmark$} 8.4\%  & \textcolor{green!50!black}{$\checkmark$} 8.7\%  & ---                                       & \textbf{1 / 4} \\
        GPT-5.4 (1.6\%)                & \textcolor{green!50!black}{$\checkmark$} 3.2\%  & \textcolor{green!50!black}{$\checkmark$} 4.4\%  & \textcolor{green!50!black}{$\checkmark$} 0.9\%  & \textcolor{green!50!black}{$\checkmark$} 2.2\%  & \textcolor{green!50!black}{$\checkmark$} 1.6\%  & \textcolor{green!50!black}{$\checkmark$} 3.5\%  & \textbf{0 / 6} \\
        DeepSeek-V3.2 (3.3\%)          & \textcolor{green!50!black}{$\checkmark$} 8.3\%  & \textcolor{green!50!black}{$\checkmark$} 3.3\%  & ---                                       & \textcolor{green!50!black}{$\checkmark$} 3.0\%  & \textcolor{green!50!black}{$\checkmark$} 3.6\%  & ---                                       & \textbf{0 / 4} \\
        Gemini 3 Flash (2.2\%)         & \textcolor{green!50!black}{$\checkmark$} 2.5\%  & \textcolor{green!50!black}{$\checkmark$} 5.7\%  & ---                                       & \textcolor{green!50!black}{$\checkmark$} 2.5\%  & ---                                       & ---                                       & \textbf{0 / 3} \\
        \midrule
        \textbf{Flagged / tested}      & 2/6 & 2/6 & 0/2 & 0/6 & 2/5 & 1/3 & \textbf{7 / 28 (25\%)} \\
        \bottomrule

    \end{tabular}
\end{table}

\partitle{Flagged Endpoints}
KBF flags \textbf{7/28} endpoints in the displayed subset, with \textbf{6/7} flags on Claude endpoints and one on GLM-4.7.
GPT-5.4, DeepSeek-V3.2, and Gemini~3 Flash are unflagged wherever available in the displayed subset.
We test KBF across deployment configurations, provider routes, and time (Appendix~\ref{sec:robustness}), but determining the causes of response-distribution inconsistencies observed at real-world API endpoints requires further independent evidence.

\begin{figure}[!htbp]
    \centering
    \captionsetup{skip=3pt}
    \captionsetup[subfigure]{font=small,skip=2pt}
    \begin{subfigure}[t]{0.485\linewidth}
        \centering
        \includegraphics[width=\linewidth]{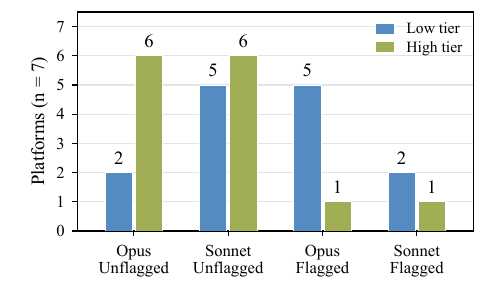}
        \caption{Audit verdicts}
        \label{fig:tier_summary_grouped}
    \end{subfigure}\hfill
    \begin{subfigure}[t]{0.485\linewidth}
        \centering
        \includegraphics[width=\linewidth]{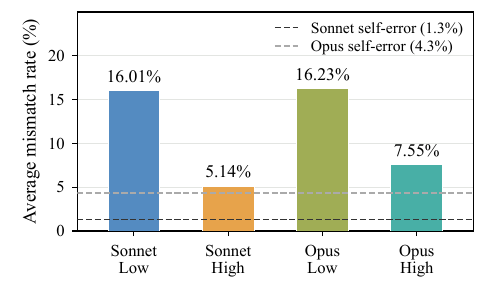}
        \caption{Average mismatch rates}
        \label{fig:tier_summary_rates}
    \end{subfigure}
    \caption{Pricing-tier comparison for Claude Opus~4.6 and Sonnet~4.6 across seven platforms. Dashed lines in (b) mark reference self-error rates.}
    \Description{Aligned panels retain all flagged and unflagged counts by model and tier, and the four original mean mismatch rates. The left panel uses blue and green for low and high tiers. The right panel distinguishes its four bars with the blue, yellow, green, and teal palette from Figure 3.}
    \label{fig:tier_comparison}
    \vspace{-5pt}
\end{figure}
\endgroup
\partitle{Tier Case Study}
Seven platforms expose Claude Opus~4.6 and Claude Sonnet~4.6 under both low and upgraded pricing tiers.
The low versus upgraded tiers are flagged on 5/7 versus 1/7 platforms for Opus and 2/7 versus 1/7 for Sonnet (Figure~\ref{fig:tier_summary_grouped}).
Average mismatch rates are about 16\% on the low tiers and 5\%--8\% on the upgraded tiers (Figure~\ref{fig:tier_summary_rates}).
For the same advertised models and platforms, low-tier responses deviate more from the reference on average.

\partitle{Audit Interpretation}
KBF is a statistical screening method whose flags indicate inconsistency with the reference response distribution on the audited probes.
\par\endgroup

\FloatBarrier
\begingroup
\setlength{\parskip}{2pt plus 0.5pt minus 0.5pt}
\vspace{-5pt}
\section{Conclusion}
\vspace{-2pt}
\label{sec:conclusion}

We study model-identity auditing for relay and reseller APIs under black-box access. We introduced \name, which fingerprints endpoints using stable factual recall near the knowledge boundary of a reference model and compares suspect endpoints with the enrolled reference using a CP binomial test. Across 16 production endpoints and eight model families, \name detected all 155 economically relevant substitutions with no false-positive rejection on same-reference controls. It remained robust under deployment-wrapper changes and detected mixed routing at economically meaningful substitution rates. In real-world shadow API audits, \name found 7/28 endpoints statistically inconsistent with their references, with inconsistencies concentrated on premium Claude endpoints.
\par\endgroup

\par\endgroup
\medskip

\subsection*{AI use statement}
Generative models serve as both the objects of study and components of the experimental protocol.
KBF uses reference LLMs to generate candidate numerical probes near their knowledge boundaries and uses Qwen3.5-9B for optional contrastive screening.
The natural-QA experiment uses GPT-4.1-mini to rewrite probes, verify semantic equivalence, and extract numerical answers from free-text responses.
Responses collected from the reference and audited endpoints form the experimental observations used for calibration and evaluation.
Section~\ref{sec:kbf} and Appendix~\ref{sec:kbf-implementation} describe this protocol, and Section~\ref{sec:setup} reports the evaluated endpoints and provider settings.

AI agents supported experimental execution, primarily by writing and debugging API request scripts, running experiments, inspecting API responses, and handling request errors and rate limits.
They also assisted with processing the experimental data, writing analysis scripts, and generating plots.

AI assistants helped with grammatical polishing and sentence-level editing of the manuscript.
They also helped draft descriptions of the experiments and their results in the experimental sections.
The authors wrote the introduction and core method sections themselves, without AI drafting assistance.
For the literature review, AI assistants were used only to search for and summarize related work.
The authors read the original papers and checked the summaries against them.

The authors developed the core research ideas and method, determined the argument and organization of the paper, and interpreted and validated the results and conclusions.
The authors reviewed all AI-assisted work, tested the code, checked the collected data and analyses, and verified the reported results.
They take full responsibility for the final manuscript, all results and conclusions, and accompanying artifacts.

\subsection*{Ethics statement}
This study audits publicly accessible LLM APIs using benign factual queries at low request volumes to limit the impact on service operation.
It involves no human participants and collects no personal information.
We de-identify the audited relay platforms and evaluate false positives and robustness under deployment variation.
An audit rejection indicates inconsistency with the reference endpoint under the tested conditions.

\subsection*{Reproducibility statement}
We describe probe construction, matching rules, calibration, and audit decisions.
Section~\ref{sec:setup} reports the evaluated endpoints and provider settings, and Appendix~\ref{app:baseline-details} documents the baseline configurations and adaptations.
Runnable scripts for probe generation and API auditing are available at \url{https://github.com/Ooo0ption/KBF}.
Running the scripts requires API access and configuration of the endpoint, model, and credentials.

\bibliographystyle{iclr2027_conference}
\bibliography{pir,refs}

\appendix
\begingroup
\renewcommand{\topfraction}{0.9}
\renewcommand{\bottomfraction}{0.85}
\renewcommand{\textfraction}{0.08}
\renewcommand{\floatpagefraction}{0.8}
\setcounter{topnumber}{3}
\setcounter{bottomnumber}{2}
\setcounter{totalnumber}{4}
\makeatletter
\setlength{\@fptop}{0pt}
\setlength{\@fpsep}{12pt plus 2pt minus 2pt}
\setlength{\@fpbot}{0pt plus 1fil}
\makeatother
\setlength{\parskip}{3pt plus 1pt minus 1pt}
\setlength{\textfloatsep}{10pt plus 2pt minus 2pt}
\setlength{\floatsep}{8pt plus 2pt minus 2pt}
\setlength{\intextsep}{8pt plus 2pt minus 2pt}
\captionsetup{skip=4pt}
\section{Detailed Method and Implementation}
\label{sec:kbf-implementation}

This appendix expands the three stages in Section~\ref{sec:kbf} and states the scope of the audit.
Algorithm~\ref{alg:kbf_enrollment} constructs the reference fingerprint through probe discovery and self-calibration.
Algorithm~\ref{alg:kbf_audit} applies the online audit to a suspect endpoint and returns \textsc{Same} or \textsc{Different}.
The implementation below uses numerical recall probes, with shared prompt rendering, batch alignment, and parsing across the three stages.

\SetKwInput{KwOracles}{Oracles}
\SetKwInput{KwInputs}{Inputs}
\SetKwInput{KwOutput}{Output}

\subsection{Audit Scope}
\label{app:problem-setting}

Relay auditing starts from a specific advertised endpoint and a queryable official reference for that endpoint.
The probes can therefore be reference-specific rather than universal.
LLMmap identifies a model from an enrolled database~\cite{llmmap}, whereas KBF compares a claimed endpoint with its enrolled reference.
KBF fingerprints the response behavior of the served API endpoint.
If the reference endpoint uses retrieval or tools, that behavior is part of the reference.
Conversely, if a suspect relay secretly augments a claimed bare-model API with retrieval, tools, or an external fact database, it is no longer serving the same endpoint under the same deployment conditions.

The economic substitution evaluation covers same-tier and lower-tier targets.
The false-positive rate is measured in same-reference trials.
Our mixed-routing analysis provides point estimates, approximate confidence intervals for known substitutes, and candidate-pool point-estimate ranges under the routing assumptions stated in Section~\ref{sec:kbf-adaptive-routing}.

\Needspace{6\baselineskip}
\subsection{Stage I: Reference Probe Generation}
\label{app:kbf-generation}

\begin{table}[!htbp]
    \centering
    \footnotesize
    \setlength{\tabcolsep}{3.5pt}
    \renewcommand{\arraystretch}{1.08}
    \caption{Example domains and matching rules.}
    \label{tab:domain_match_examples}
    \begin{tabularx}{\linewidth}{>{\raggedright\arraybackslash}X p{1.65cm} p{2.9cm}}
        \toprule
        \textbf{Domain} & \textbf{Valid range} & \textbf{$\mathsf{match}_d(a,\hat{a})$} \\
        \midrule
        Material boiling point & $[-300,600]$ & $\lvert a-\hat{a}\rvert \leq 3$ $^\circ$C \\
        Astronomy facts & $[0,10^{15}]$ & relative error $\leq 5\%$ \\
        Chromosome count & $[1,2000]$ & exact agreement \\
        Programming release year & $[1950,2030]$ & exact year agreement \\
        Drug half-life & $[0.01,5000]$ & relative error $\leq 10\%$ \\
        \bottomrule
    \end{tabularx}
\end{table}
\partitle{Domain library}
The auditor searches each domain independently for boundary probes and retains those with stable reference completions.
Our implementation uses 15 domains, covering chemistry, biology, astronomy, math (e.g., OEIS sequences~\cite{oeis}), programming, and cryptography.
Each domain specifies a prompt template, a valid numerical range, a difficulty schedule, and a comparison rule $\mathsf{match}_d$.
Table~\ref{tab:domain_match_examples} gives examples.
Auditors can add or replace domains as long as the facts are numerical, parseable, and produce stable boundary recall.

\partitle{Candidate proposal}
Each search round asks the reference endpoint to propose structured candidate records as \texttt{name | value} pairs.
The implementation specifies a topic and a difficulty setting, and appends an exclusion list of previously proposed names.
The exclusion list reduces repeated proposals during the search.
Difficulty progression is an implementation choice. The component ablations in Appendix~\ref{app:signal-ablation} hold difficulty fixed while examining specialist candidate generation and the two screening components.

\partitle{Recall prompts}
KBF parses the proposed records, removes duplicate names, rejects values outside the domain range, and renders the remaining records into short audit prompts $q_i$.
Our implementation uses a cloze template for each domain, so a boiling-point record becomes ``The boiling point of $x$ at 1 atm is \_\_ $^\circ$C.''
The endpoint completes the sentence, leaving a numerical slot that can be recovered under the same rules during consistency checks, self-calibration, and auditing.
We use two short, task-oriented prompts and avoid complex behavioral instructions.
The system prompt requests adherence to the task, and the user prompt asks for factual recall with one numbered answer per probe.

The candidate-generation, system, and user prompts can all be modified, provided they preserve the probe semantics, yield parseable answers, and support stable reference recall.
The auditor applies the revised wording consistently across reference checks, self-calibration, and the corresponding online audit.
This flexibility reduces dependence on a fixed template.
Appendix~\ref{app:probe-formats} discusses relay recognition and evaluates four alternative probe formats.

\partitle{Screening and stopping}
After the first reference query under $c_{\mathrm{audit}}$, KBF can query a small contrast endpoint (default Qwen3.5-9B) on the same prompt.
If it matches the provisional reference value under $\mathsf{match}_d$, the candidate is dropped before the full consistency checks.
This saves reference calls and biases the retained set toward candidates better separated from cheap substitutes.
The screen is a probe-selection heuristic, so the online audit still compares the suspect endpoint only against the enrolled reference behavior.
We disable it for reference models where it removes too much of the usable probe set.

Reference-consistency checks re-query the surviving candidates, including at $\mathrm{temp}{=}0.7$ to filter out unstable probes.
A candidate is enrolled only when the responses are valid and agree under $\mathsf{match}_d$.
KBF stores the prompt, domain, consensus value, and comparison rule in $P_{\mathrm{ref}}$.
After each round, it appends the number of surviving probes $|S_{d,t}|$ to the stable-count history $H_d$.
The predicate $\mathrm{StopDomainCriteria}(H_d)$ fires after two consecutive zero-yield rounds once the domain has enrolled at least five stable probes.
The search also ends when it exhausts the difficulty schedule supplied to Algorithm~\ref{alg:kbf_enrollment}.

\subsection{Stage II: Reference Self-Calibration}
\label{app:kbf-calibration}

\partitle{Self-test and stored fingerprint}
The auditor fixes $P_{\mathrm{ref}}$ and re-queries the reference endpoint under the same configuration used for online audits, with $\mathrm{temp}{=}0$.
Missing, unparseable, and out-of-range answers are excluded from scoring. The self-test denominator $n_{\mathrm{self}}$ counts valid responses, and $k_{\mathrm{self}}$ counts nonmatching responses among them.
The auditor sets $p_0=\mathrm{CP}_{99}(k_{\mathrm{self}},n_{\mathrm{self}})$ and stores $F_{\mathrm{ref}}=(P_{\mathrm{ref}},k_{\mathrm{self}},p_0)$.
Algorithm~\ref{alg:kbf_enrollment} gives both enrollment stages, including the search state, stopping rule, and stored output.
The implementation also stores raw consensus responses and per-probe self-test outcomes for diagnostics and the matched-probe routing test in Section~\ref{sec:kbf-routing-overview}.

\begin{algorithm}[t]
\caption{KBF Reference Fingerprint Construction}
\label{alg:kbf_enrollment}
\DontPrintSemicolon
\KwOracles{Reference endpoint $\mathcal{O}_{\mathrm{ref}}$}
\KwInputs{Search space $\mathcal{D}$ with domains, audit prompt templates, and match rules; difficulty schedule $\mathcal{T}$; reference-consistency checks $\mathcal{C}_{\mathrm{cons}}$; audit configuration $c_{\mathrm{audit}}$; self-calibration confidence $\gamma$}
\KwOutput{Reference fingerprint $F_{\mathrm{ref}}=(P_{\mathrm{ref}},k_{\mathrm{self}},p_0)$}
\BlankLine
\textbf{\textsc{Stage I: Probe Discovery}}\;
\Indp
$P_{\mathrm{ref}} \leftarrow \emptyset$\;
\ForEach{domain $d \in \mathcal{D}$}{
    $\mathsf{match}_d \leftarrow$ domain matching rule\;
    $H_d \leftarrow$ empty stable-count history for domain $d$\;
    \ForEach{search round $t \in \mathcal{T}$}{
        $P_{d,t} \leftarrow$ candidate probes proposed by $\mathcal{O}_{\mathrm{ref}}$ for domain $d$ at difficulty $t$\;
        $S_{d,t} \leftarrow \emptyset$\;
        \ForEach{candidate probe $q_i \in P_{d,t}$}{
            $A_i \leftarrow$ responses from $\mathcal{O}_{\mathrm{ref}}$ to $q_i$ under check configuration $c \in \mathcal{C}_{\mathrm{cons}}$\;
            \If{$A_i$ is valid and stable under $\mathsf{match}_{d}$}{
                $a_i \leftarrow \mathrm{Consensus}(A_i,\mathsf{match}_{d})$\;
                $S_{d,t} \leftarrow S_{d,t} \cup \{(q_i,d,a_i,\mathsf{match}_{d})\}$\;
            }
        }
        $H_d \leftarrow \mathrm{Append}(H_d, |S_{d,t}|)$\;
        $P_{\mathrm{ref}} \leftarrow P_{\mathrm{ref}} \cup S_{d,t}$\;
        \If{$\mathrm{StopDomainCriteria}(H_d)$}{
            \textbf{break}\;
        }
    }
}
\Indm
\BlankLine
\textbf{\textsc{Stage II: Self-Calibration}}\;
\Indp
$k_{\mathrm{self}} \leftarrow 0$; $n_{\mathrm{self}} \leftarrow 0$\;
\ForEach{$(q_i,d_i,a_i,\mathsf{match}_{d_i}) \in P_{\mathrm{ref}}$}{
    $\tilde{a}_i \leftarrow \mathcal{O}_{\mathrm{ref}}(q_i,c_{\mathrm{audit}})$\;
    \If{$\tilde{a}_i$ is domain-valid}{
        $n_{\mathrm{self}} \leftarrow n_{\mathrm{self}}+1$\;
        \If{$\mathsf{match}_{d_i}(a_i,\tilde{a}_i)=0$}{
            $k_{\mathrm{self}} \leftarrow k_{\mathrm{self}}+1$\;
        }
    }
}
$p_0 \leftarrow \mathrm{CP}_{\gamma}(k_{\mathrm{self}}, n_{\mathrm{self}})$\;
\Indm
\BlankLine
\Return{$(P_{\mathrm{ref}},k_{\mathrm{self}},p_0)$}\;
\end{algorithm}

\partitle{Null calibration and interpretation}
Self-calibration estimates ordinary reference-side noise under the deployed interface, including nondeterministic generation, quantization, cache effects, provider-side wrappers, and other serving details outside the auditor's control.
Under the binomial model, each scored probe is treated as an independent discrepancy trial with the same unknown reference error rate $p$.
The $\mathrm{CP}_{\gamma}$ value is a $\gamma$-level one-sided upper confidence bound on this rate.
For $0\leq k<n$, it is given by $\mathrm{CP}_{\gamma}(k,n)=\mathrm{Beta}^{-1}(\gamma;\,k+1,\,n-k)$, with $\mathrm{CP}_{\gamma}(n,n)=1$.
If the true reference error rate were larger than this bound, observing $k$ or fewer discrepancies would fall in the lower $1-\gamma$ tail.
A stable reference endpoint with many enrolled probes receives a tight null bound, while a noisier endpoint receives a wider bound.

Calibration and auditing use batch validation and retries to handle extensive parsing failures and detected answer misalignment.
Responses that remain invalid are excluded from both discrepancy counts and denominators.
The binomial model applies to these final scored responses, with independence as a modeling assumption.
Shared request context or endpoint state can still affect several valid answers together, so the reported $p$-values should be interpreted under this assumption.

We report the resulting significance together with independent same-reference controls and robustness results, which provide empirical evidence under the tested deployment conditions.

\subsection{Stage III: Suspect Endpoint Audit}
\label{app:kbf-audit}

\partitle{Batched queries}
The auditor loads $F_{\mathrm{ref}}$ and sends the enrolled prompts to the suspect endpoint under $c_{\mathrm{audit}}$.
Oracle calls are issued in domain-homogeneous batches of ten probes, so a batch shares one valid range and one match rule.
The recall prompt requests numbered answers, letting the parser map each value back to its probe when an endpoint skips a line, inserts commentary, or reorders answers.
The same batching and parsing procedure is used for reference checks and self-calibration.

\partitle{Response parsing}
KBF uses one numerical parser for $A_i$, $\tilde{a}_i$, and $\hat{a}_i$.
It normalizes each response, aligns the numbered answers with their probe slots, and extracts a final value, rejecting anything outside the valid range.
A recoverable slot is kept when alignment is clear, and a formatting failure or ambiguous alignment triggers a retry.
A probe succeeds only when the recovered value is domain-valid and matches the stored consensus.
Missing, unparseable, and out-of-range answers are excluded from both the discrepancy count and the audit denominator $n$. Valid but nonmatching answers count as discrepancies.

\partitle{Decision and diagnostics}
Algorithm~\ref{alg:kbf_audit} counts these discrepancies and applies the one-sided binomial test with the stored null bound $p_0$ at $\alpha=0.05$.
It reports \textsc{Different} when the observed count exceeds what calibrated reference self-noise can explain, and \textsc{Same} otherwise.
The decision is endpoint-level evidence under the audited interface.
The implementation records McNemar's paired test~\cite{mcnemar} against the stored self-test outcomes as an auxiliary diagnostic for quantization and provider-comparison analyses.
The primary relay-audit decision uses the CP-calibrated binomial rule.

\begin{algorithm}[t]
\caption{KBF Endpoint Audit}
\label{alg:kbf_audit}
\DontPrintSemicolon
\KwOracles{Suspect endpoint $\mathcal{O}_{\mathrm{sus}}$}
\KwInputs{Reference fingerprint $F_{\mathrm{ref}}=(P_{\mathrm{ref}},k_{\mathrm{self}},p_0)$; audit configuration $c_{\mathrm{audit}}$; significance level $\alpha$}
\KwOutput{\textsc{Same} or \textsc{Different}}
$n \leftarrow 0$; $k \leftarrow 0$\;
\ForEach{$(q_i,d_i,a_i,\mathsf{match}_{d_i}) \in P_{\mathrm{ref}}$}{
    $\hat{a}_i \leftarrow \mathcal{O}_{\mathrm{sus}}(q_i,c_{\mathrm{audit}})$\;
    \If{$\hat{a}_i$ is domain-valid}{
        $n \leftarrow n+1$\;
        \If{$\mathsf{match}_{d_i}(a_i,\hat{a}_i)=0$}{
            $k \leftarrow k+1$\;
        }
    }
}
$p \leftarrow \Pr[X \geq k \mid X \sim \mathrm{Binomial}(n,p_0)]$\;
\eIf{$p < \alpha$}{
    \Return{\textsc{Different}}\;
}{
    \Return{\textsc{Same}}\;
}
\end{algorithm}

\Needspace{6\baselineskip}
\subsection{Detection with Boundary Probes}
\label{app:boundary-characterization}

\partitle{Experimental setup}
E16 uses Claude Sonnet~4.6, GPT-5.4, DeepSeek-V3.2, and GPT-4.1-mini as references.
For each reference, we generate candidates without early stopping across 15 domains and eight rounds per domain under each setting in Appendix~\ref{app:signal-ablation-setup}, then freeze 800 probes per setting.
Of these 1,600 probes, 378--658 belong to $K$, passing reference-consistency checks while disagreeing with the construction contrast, Qwen3.5-9B.
We evaluate this subset using separate calibration, same-reference, and cross-model response batches, with the other three references serving as targets.
We exclude unparseable responses, estimate a $\mathrm{CP}_{99}$ null bound from calibration, and apply a one-sided binomial test at $\alpha=0.01$.

\partitle{Detection results}
Using only $K$ probes yields $\mathrm{FP}=0/4$ on the same-reference controls and $\mathrm{TP}=12/12$ across the directed cross-model pairs.
Thus, boundary probes alone support calibrated detection in this four-model panel.

\Needspace{6\baselineskip}
\section{Probe Formats}
\label{app:probe-formats}
\label{app:relay-recognition}

\partitle{Probe formats}
We construct four variants of the original numerical cloze probes by changing the answer type, question wording, task wrapper, or request grouping.
Table~\ref{tab:probe-format-examples} compares their request features and gives an example of each format.

\begin{enumerate}[leftmargin=*,itemsep=3pt,topsep=3pt]
    \item Entity cloze revises the candidate-generation rules to produce non-numerical probes whose answers are entity names.
It retains the fill-in-the-blank format and compares answers by normalized string matching.

    \item Numerical QA generates direct questions with numerical answers, replacing the fill-in-the-blank format while retaining the fixed task wrapper.

    \item Natural batch QA uses GPT-4.1-mini to rewrite the numerical questions and verify semantic equivalence, then removes the fixed task wrapper and answer-format instructions.
Questions remain batched, and GPT-4.1-mini extracts numerical answers from free-text replies.

    \item Single-question QA sends the same rewritten questions one per request, removing the batch structure and brief-answer opener while retaining free-text answer extraction.
\end{enumerate}

\begin{table}[!htbp]
    \centering
    \footnotesize
    \renewcommand{\arraystretch}{1.15}
    \setlength{\tabcolsep}{4pt}
    \caption{KBF probe formats and detection results. Examples use GPT-5.4 reference probes. $\bullet$ = retained, $\circ$ = removed. The features are numerical answers (Number), blanks (Cloze), a fixed task prompt and answer-format instruction (Wrapper), and ten-question requests (Batch). TP counts detected substitutions and FP counts same-reference false positives at $\alpha=0.05$. Numerical cloze uses the full 16-model evaluation in Section~\ref{sec:detection}, while the four variants use the five-model experiment below. Answers for the two wrapper-free variants are extracted by GPT-4.1-mini.}
    \label{tab:probe-format-examples}
    \begin{tabularx}{\linewidth}{@{}>{\raggedright\arraybackslash}p{1.55cm}@{\hspace{3pt}}c@{\hspace{3pt}}c@{\hspace{3pt}}c@{\hspace{3pt}}c@{\hspace{5pt}}>{\raggedright\arraybackslash}X@{\hspace{5pt}}>{\raggedright\arraybackslash}p{1.5cm}@{}}
        \toprule
        \textbf{Format} & \rotatebox[origin=l]{45}{\textbf{Number}} & \rotatebox[origin=l]{45}{\textbf{Cloze}} & \rotatebox[origin=l]{45}{\textbf{Wrapper}} & \rotatebox[origin=l]{45}{\textbf{Batch}} & \textbf{Example query and stored answer} & \textbf{Detection} \\
        \midrule
        Numerical cloze$^{*}$ & $\bullet$ & $\bullet$ & $\bullet$ & $\bullet$ & The boiling point of hexamethyldisilazane at 1 atm is \_\_\_$^\circ$C.\newline Answer: 126. & TP 155/155\newline FP 0/16 \\
        \addlinespace[5pt]
        Entity cloze & $\circ$ & $\bullet$ & $\bullet$ & $\bullet$ & The author of ``Fathers of Nations'', a prescribed set text for the KCSE examination, is \_\_\_.\newline Answer: Paul B. Vitta. & TP 10/10\newline FP 0/5 \\
        \addlinespace[5pt]
        Numerical QA & $\bullet$ & $\circ$ & $\bullet$ & $\bullet$ & At 1 atmosphere pressure, what is the boiling point in degrees Celsius of hexamethyldisilazane, $[(\mathrm{CH}_3)_3\mathrm{Si}]_2\mathrm{NH}$?\newline Answer: 126. & TP 10/10\newline FP 0/5 \\
        \addlinespace[5pt]
        Natural batch QA & $\bullet$ & $\circ$ & $\circ$ & $\bullet$ & Request opens with ``A few quick questions, brief answers are fine:'' followed by ten rewritten questions. Item~1: What is the boiling point in degrees Celsius of hexamethyldisilazane, $[(\mathrm{CH}_3)_3\mathrm{Si}]_2\mathrm{NH}$, at 1 atmosphere pressure?\newline Answer: 126, extracted from the free-text reply. & TP 10/10\newline FP 0/5 \\
        \addlinespace[5pt]
        Single-question QA & $\bullet$ & $\circ$ & $\circ$ & $\circ$ & What is the boiling point in degrees Celsius of trimethylsilyl bromide, $(\mathrm{CH}_3)_3\mathrm{SiBr}$, at 1 atmosphere pressure?\newline Answer: 79.5, extracted from the reply ``about 79--80\,$^\circ$C''. & TP 10/10\newline FP 0/5 \\
        \bottomrule
    \end{tabularx}
    \par\smallskip
    {\footnotesize $^{*}$Original numerical cloze probes from the 16-model evaluation in Section~\ref{sec:detection}.}
\end{table}

\partitle{Evaluation protocol}
We evaluate all four variants on Claude Sonnet~4.6, GPT-5.4, GLM-4.7, GPT-4.1-mini, and Qwen3.5-27B.
Each variant follows the same enrollment and audit procedure as the main experiment in Section~\ref{sec:detection}, using its own probe format throughout.

\partitle{Detection results}
Table~\ref{tab:probe-format-examples} shows that all four variants detect all ten tested downgrades with no false positives in five same-reference trials per variant.

\partitle{Further variants and extensions}
The core of KBF is the factual question and the stable reference answer it elicits near the knowledge boundary.
This signal supports variation in how a question is phrased, presented, and answered, as the four experiments demonstrate.
The construction can extend to additional factual domains, other short-answer types, application-specific wording, mixed-domain requests, and probes embedded in ordinary workloads.
Each extension requires a suitable answer-extraction and matching rule, reference-consistency checks in the intended format, and fresh self-calibration.
Section~\ref{sec:config-robustness} provides separate evidence for the tested prompt and decoding changes.

\section{Supplement to Detection Accuracy and Cost}
\label{app:detection-extended}

Section~\ref{sec:detection} reports the 155 economically relevant directions of the 16$\times$16 matrix.
This appendix adds two results that the main text omits.
The first is the complete matrix, including the 85 upgrade directions that Figure~\ref{fig:heatmap_detection} leaves blank.
The second is a pool of probe sets enrolled after the main evaluation was frozen, together with independent same-model retests.
All reported audits use the decision rule of Section~\ref{sec:construction}. The $\mathrm{CP}_{99}$ bound on the calibration self-test sets $p_0$, and a one-sided binomial test at $\alpha=0.05$ decides each audit.
We also provide per-model enrollment costs.

\subsection{Upgrade Directions}
\label{app:upgrade-directions}

\partitle{Why upgrades are outside the threat model}
An upgrade direction is a pair $(M_r, M_t)$ in which the served target $M_t$ is more expensive than the claimed reference $M_r$, for instance an endpoint that advertises GPT-4.1-nano but serves GPT-5.4.
Such substitutions may lack an economic incentive and are excluded from the TPR in Section~\ref{sec:detection}.
Every pairwise audit was nevertheless run in both directions, and Figure~\ref{fig:e1_full_matrix} shows all 240 off-diagonal cells.

\begin{figure}[!htbp]
    \centering
    \includegraphics[width=0.92\textwidth]{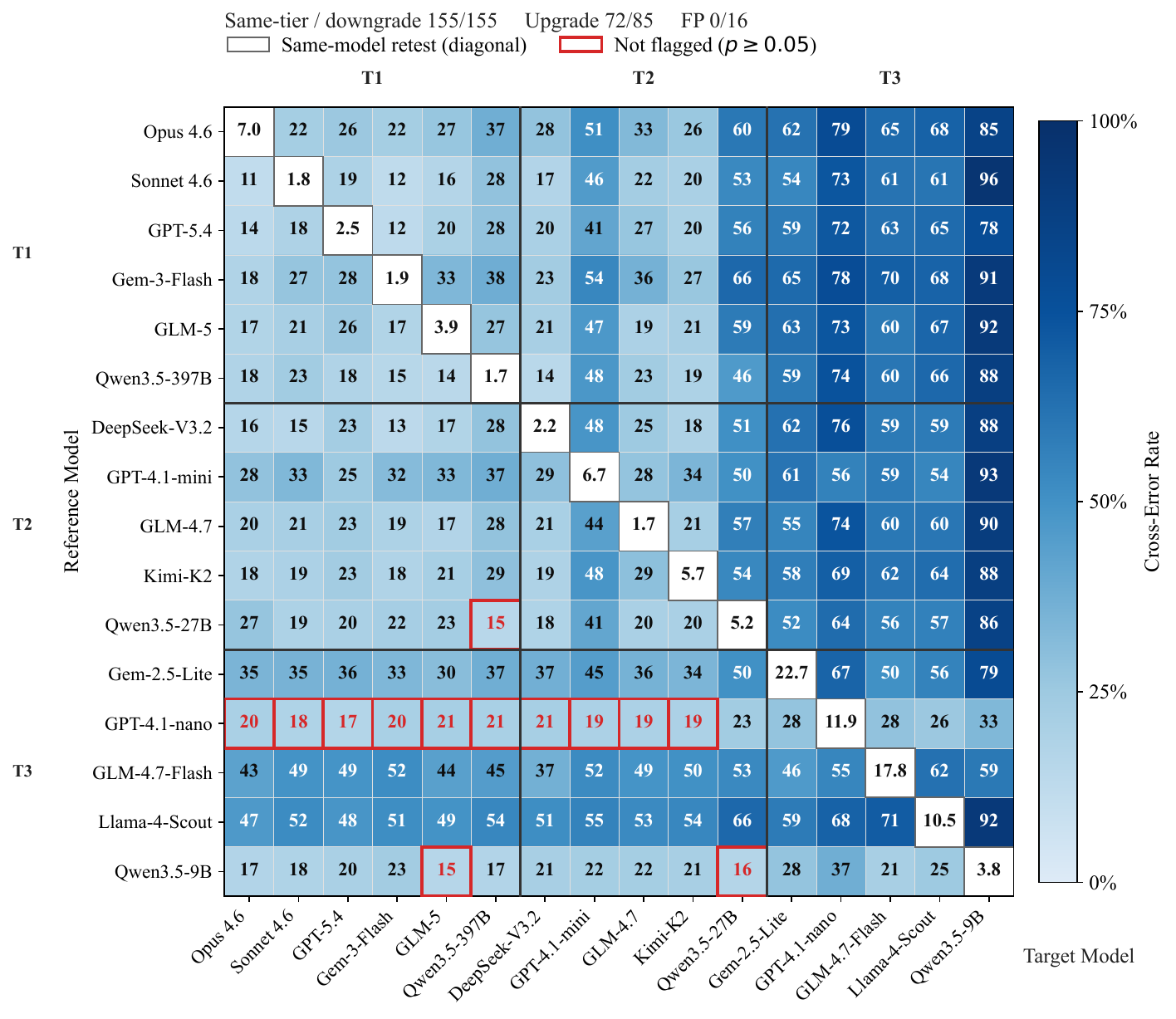}
    \caption{The complete 16$\times$16 detection matrix, including the 85 upgrade directions that Figure~\ref{fig:heatmap_detection} leaves blank. Each cell is the disagreement rate of the served target (column) on the reference probe set (row). Diagonal cells show the independent same-model retest of each reference, scored against the bound set by its calibration self-test. Red outlines mark the 13 pairs that the test does not flag at $\alpha=0.05$; all of them are upgrade directions.}
    \Description{A 16 by 16 heatmap with every off-diagonal cell filled. Thirteen cells, all in the upgrade direction and concentrated in the GPT-4.1-nano row, are outlined in red to mark pairs that were not flagged.}
    \label{fig:e1_full_matrix}
\end{figure}

\partitle{Results}
KBF flags 72 of the 85 upgrade directions, so 227 of the 240 measured substitutions are detected overall.
The 13 misses are outlined in red in Figure~\ref{fig:e1_full_matrix}.
They come from three references only: GPT-4.1-nano accounts for 10, Qwen3.5-9B for 2, and Qwen3.5-27B for 1.
In every case the reverse direction is flagged with a wide margin.
The Qwen3.5-397B-A17B probe set flags Qwen3.5-27B at a 46\% disagreement rate, while the Qwen3.5-27B probe set reads Qwen3.5-397B-A17B at 15\% and does not reject.
The GPT-4.1-nano probe set reads every T1 target between 17\% and 21\%, whereas every T1 probe set reads GPT-4.1-nano between 72\% and 79\%.

\partitle{Directional asymmetry}
Probes target the knowledge boundary of the reference model. A stronger target can reproduce many of its correct answers, whereas a weaker target fails on more of its boundary facts.
All 13 unflagged upgrade pairs involve references with at most 115 retained probes and a null bound $p_0\geq10.7\%$, reaching 15.3\% for GPT-4.1-nano.

\subsection{Frontier Probe Sets}
\label{app:frontier-sets}

\partitle{New probe sets}
The endpoint pool of Table~\ref{tab:models} was frozen in March 2026.
Between 10 and 18 September 2026 we ran the released enrollment pipeline, unchanged, against 28 endpoints: 13 models released after the freeze and 15 of the original 16 (GLM-4.7-Flash was not re-enrolled).
Table~\ref{tab:frontier_sets} lists them with the pinned provider, the OpenRouter list price recorded on 19 September 2026, the retained probe count, and the enrollment self-disagreement rate that sets $p_0$.
T1, T2, and T3 denote flagship, mid-range, and budget models, respectively. Within T1, T1-1 denotes the latest flagship generation as of September 2026, and T1-2 denotes the generation that was latest in March 2026. Generation and product tier are separate attributes, so a newer mid-range model such as GPT-5.6 Luna remains in T2.
Budget-tier references were enrolled without contrastive screening, because the default contrast model is itself a budget model.
Probe counts range from 149 to 557, and the median calibration self-disagreement rate is 6.7\%.
None of the 28 retests is rejected, so this pool adds 0/28 false positives to the 0/16 of Section~\ref{sec:fp-control}.

\begin{table}[!htbp]
    \centering
    \footnotesize
    \setlength{\tabcolsep}{2pt}
    \caption{The 28 probe sets enrolled in September 2026. $^{\bullet}$ marks the 13 models released after the March 2026 freeze of Table~\ref{tab:models}. T1-1 and T1-2 denote the latest flagship generations as of September and March 2026, respectively. \emph{Model Provider} is the OpenRouter inference provider pinned for that enrollment. \emph{Price}$^{*}$ reports input- and output-token costs in USD per million tokens. \emph{Self-Err} is the enrollment self-disagreement rate, whose $\mathrm{CP}_{99}$ upper bound is the null bound $p_0$. $^{*}$Prices were recorded from OpenRouter in September 2026.}
    \label{tab:frontier_sets}
    \begin{tabular}{@{}llll | rr | cc@{}}
        \noalign{\hrule height 1.5pt}
        \multirow{2}{*}{\textbf{Tier}} & \multirow{2}{*}{\textbf{Model}} & \multirow{2}{*}{\textbf{Model Family}} & \multirow{2}{*}{\textbf{Model Provider}} &
        \multicolumn{2}{c|}{\textbf{Price$^{*}$ (\$/M Tokens)}} &
        \multirow{2}{*}{\textbf{\#Probes}} & \multirow{2}{*}{\textbf{Self-Err}} \\
        \cline{5-6}
        & & & & \textbf{Input} & \textbf{Output} & & \\
        \hline
        T1-1 & Claude Fable 5.1$^{\bullet}$ & Anthropic & Anthropic & 10.00 & 50.00 & 453 & 2.6\% \\
T1-1 & GPT-6 Astra$^{\bullet}$ & OpenAI & OpenAI & 10.00 & 50.00 & 538 & 1.5\% \\
T1-1 & Claude Opus 5$^{\bullet}$ & Anthropic & Claude on AWS &  5.00 & 25.00 & 414 & 1.7\% \\
T1-1 & GPT-5.6 Sol$^{\bullet}$ & OpenAI & OpenAI &  2.00 & 10.00 & 426 & 6.8\% \\
T1-1 & Kimi-K3$^{\bullet}$ & Moonshot & Moonshot AI &  1.70 &  8.50 & 424 & 8.5\% \\
T1-1 & GLM-5.3$^{\bullet}$ & Z.AI & Z.AI &  0.91 &  2.86 & 372 & 7.3\% \\
\hline
T1-2 & Claude Opus 4.6 & Anthropic & Amazon Bedrock &  5.00 & 25.00 & 445 & 5.2\% \\
T1-2 & Claude Sonnet 4.6 & Anthropic & Google &  3.00 & 15.00 & 436 & 6.4\% \\
T1-2 & GPT-5.4 & OpenAI & OpenAI &  2.50 & 15.00 & 531 & 7.7\% \\
T1-2 & GLM-5 & Z.AI & Z.AI &  0.60 &  1.92 & 332 & 6.6\% \\
T1-2 & Qwen3.5-397B-A17B & Alibaba & Alibaba &  0.55 &  3.50 & 254 & 6.3\% \\
T1-2 & Gemini 3 Flash & Google & Google &  0.50 &  3.00 & 409 & 5.1\% \\
\hline
T2 & Hunyuan 4 Preview$^{\bullet}$ & Tencent & Tencent &  0.83 &  2.50 & 538 & 5.8\% \\
T2 & GPT-5.4-mini$^{\bullet}$ & OpenAI & OpenAI &  0.75 &  4.50 & 194 & 13.4\% \\
T2 & Kimi-K2-0905 & Moonshot & Novita &  0.60 &  2.50 & 363 & 7.8\% \\
T2 & DeepSeek-V4-Pro$^{\bullet}$ & DeepSeek & Novita &  0.58 &  1.73 & 557 & 6.1\% \\
T2 & Seed 2.1 Turbo$^{\bullet}$ & ByteDance & Seed &  0.50 &  2.50 & 400 & 6.0\% \\
T2 & GLM-4.7 & Z.AI & Z.AI &  0.40 &  1.75 & 430 & 7.7\% \\
T2 & GPT-4.1-mini & OpenAI & OpenAI &  0.40 &  1.60 & 265 & 6.4\% \\
T2 & DeepSeek-V3.2 & DeepSeek & Google &  0.27 &  0.40 & 380 & 6.1\% \\
T2 & GPT-5.6 Luna$^{\bullet}$ & OpenAI & OpenAI &  0.20 &  1.20 & 305 & 7.6\% \\
T2 & Qwen3.5-27B & Alibaba & Alibaba &  0.20 &  1.56 & 149 & 6.7\% \\
\hline
T3 & MiniMax M3$^{\bullet}$ & MiniMax & Minimax &  0.30 &  1.20 & 311 & 8.1\% \\
T3 & DeepSeek-V4.1-Flash$^{\bullet}$ & DeepSeek & Novita &  0.15 &  0.60 & 363 & 8.0\% \\
T3 & Gemini 2.5 Flash Lite & Google & Google &  0.10 &  0.40 & 319 & 10.7\% \\
T3 & GPT-4.1-nano & OpenAI & OpenAI &  0.10 &  0.40 & 248 & 10.1\% \\
T3 & LLaMA-4-Scout & Meta & Novita &  0.10 &  0.30 & 231 & 15.2\% \\
T3 & Qwen3.5-9B & Alibaba & Together &  0.10 &  0.15 & 240 & 5.0\%
\\
        \noalign{\hrule height 1.5pt}
    \end{tabular}
\end{table}

\Needspace{6\baselineskip}
\subsection{Reference Enrollment Cost}
\label{app:enrollment-cost}

Table~\ref{tab:t1_probe_generation_cost} reports a separate May 2026 cost run for the six T1 references.
This run measures enrollment cost, so its probe counts differ from those in Table~\ref{tab:models}.
It retained 3,696 probes for \$7.07 using 1,936,012 tokens, with most of the expense coming from Claude Sonnet~4.6, Claude Opus~4.6, and GPT-5.4.
The enrolled fingerprints can be reused across audits of endpoints claiming the same reference version.

\begin{table}[!htbp]
    \centering
    \small
    \setlength{\tabcolsep}{4pt}
    \caption{One-time T1 reference-enrollment cost.}
    \label{tab:t1_probe_generation_cost}
    \begin{tabular}{lccc}
        \toprule
        \textbf{Model} & \textbf{Probes} & \textbf{Cost (USD)} & \textbf{Tokens} \\
        \midrule
        Claude Sonnet 4.6      & 664 & \$2.17 & 346,316 \\
        Claude Opus 4.6        & 724 & \$2.06 & 280,718 \\
        GPT-5.4                & 531 & \$1.59 & 297,718 \\
        Gemini 3 Flash         & 716 & \$0.44 & 348,041 \\
        GLM-5                  & 613 & \$0.50 & 345,303 \\
        Qwen3.5-397B-A17B      & 448 & \$0.31 & 317,916 \\
        \midrule
        \textbf{Total}         & \textbf{3,696} & \textbf{\$7.07} & \textbf{1,936,012} \\
        \bottomrule
        \multicolumn{4}{l}{\footnotesize Re-measured in May 2026, for cost assessment only.} \\
    \end{tabular}
\end{table}

\section{Additional Robustness Analysis}
\label{sec:robustness}

We assess KBF robustness under adversarial prompts, identity spoofing, quantization, backend-provider variation, temporal drift, threshold choice, agent interfaces, and deployment wrappers across the full model pool. Table~\ref{tab:robustness_summary} summarizes the results.

\begin{mdframed}[linewidth=0.7pt,linecolor=black!65,backgroundcolor=black!3,
    innerleftmargin=9pt,innerrightmargin=9pt,innertopmargin=7pt,innerbottommargin=7pt,
    skipabove=7pt,skipbelow=9pt,nobreak=true]
\partitle{Key takeaway}
Most tested settings preserve the audit decision.
Two adversarial prompt settings cause false rejections, one that explicitly requests approximate answers and one small probe set with high baseline self-error. Temporal retests identify drift in a small subset of endpoints.
\end{mdframed}

\begin{table}[!htb]
    \centering
    \small
    \renewcommand{\arraystretch}{1.18}
    \setlength{\tabcolsep}{4pt}
    \caption{Additional robustness tests. Each row summarizes the test coverage, audit outcome, and practical conclusion.}
    \label{tab:robustness_summary}
    \begin{tabularx}{\linewidth}{@{}l>{\raggedright\arraybackslash}p{1.85cm}>{\raggedright\arraybackslash}p{1.9cm}>{\raggedright\arraybackslash}p{2.2cm}>{\raggedright\arraybackslash}X@{}}
        \toprule
        \textbf{ID} & \textbf{Test} & \textbf{Coverage} & \textbf{Audit outcome} & \textbf{Conclusion} \\
        \midrule
        \textbf{R1} & Adversarial prompts & 3 models, 5 settings & \textbf{13/15} pass & KBF tolerates most tested prompt changes, with failures under precision loss or a small, noisy probe set. \\
        \addlinespace[3pt]
        \textbf{R2} & Identity spoofing & 15 substitutes & \textbf{0/15} false positives & Claimed identities do not override the measured fingerprint in these tests. \\
        \addlinespace[3pt]
        \textbf{R3} & Quantization & 9 backends & \textbf{0/9} flagged & The tested quantized backends preserve the audit decision despite detectable differences in individual probe responses. \\
        \addlinespace[3pt]
        \textbf{R4} & Backend providers & 5 routes, Sonnet 4.6 & KBF \textbf{0/5} rejected, IRIS \textbf{1/5} & KBF tolerates the tested provider variation, while IRIS falsely rejects one route. \\
        \addlinespace[3pt]
        \textbf{R5} & Temporal drift & 5 snapshots, then 11 endpoints retested on day 179 & \textbf{16/16} stable at week 7, \textbf{10/11} at follow-up & Observed stability supports probe reuse over several weeks, with later drift motivating periodic refresh. \\
        \addlinespace[3pt]
        \textbf{R6} & Threshold choice & 10 margins & \textbf{0/16} false positives at margins of 4\%--10\% & Detection tolerates a broad margin range, with 100\% TPR at 1--6\% and 99.4\% at 7\%. \\
        \addlinespace[3pt]
        \textbf{R7} & Agent interface & 3 CLIs, 4 configurations & \textbf{0/4} false positives & KBF supports stable model auditing through the tested Claude Code, OpenClaw, and Codex interfaces. \\
        \addlinespace[3pt]
        \textbf{R8} & Extended KBF stress test & 16 models, 5 configurations & \textbf{2/80} false positives & All T1 and T2 models remain stable. GLM-4.7-Flash accounts for both false positives. \\
        \bottomrule
    \end{tabularx}
\end{table}

\partitle{R1. Adversarial prompts}
A dishonest relay may target the audit surface directly.
We test two variants: adversarial prompt settings that make numerical extraction harder, and identity-spoofing prompts that ask a substitute model to present itself as the reference.

For prompt-level stress, we subject three models (DeepSeek-V3.2, Kimi-K2-0905, GPT-4.1-nano) to five settings: temperature 1.0, an ``approximate answers'' system prompt, a non-English system prompt with English probes, 2000-character padding, and contradictory few-shot examples.
\textbf{13 of 15 trials pass}; both failures have clear root causes.
Kimi-K2-0905 reaches a 20.0\% mismatch rate under the ``approximate answers'' prompt, which directly instructs the model to abandon numerical precision.
GPT-4.1-nano reaches 18.5\% at temperature 1.0; this case combines a small probe set (109 probes) with elevated baseline self-error, leaving limited statistical separation.
DeepSeek-V3.2 passes all five settings, with a maximum error shift of 3.3\%.

\partitle{R2. Identity spoofing}
For identity spoofing, we prepend to each of the other 15 models a system prompt asserting ``You are Claude Opus~4.6'' at temperature 0.0, with the impersonated reference itself excluded.
Under the same $\mathrm{CP}_{99}$ binomial test, \textbf{all 15 endpoints stay consistent with their own baseline} (0/15 FP).
These results support robustness to the tested identity prompt when comparing each endpoint with its own reference.

\partitle{R3. Quantization robustness}
Providers may serve quantized variants (fp8, fp4) under the same model name.
This test measures whether such backend choices change KBF's same-endpoint decision.
We route GLM-4.7 queries through five OpenRouter providers spanning fp8 and fp4, anchored to the full-precision Z.AI reference, and DeepSeek-V3.2 queries through four providers anchored to the full-precision Google reference.
\textbf{None of the nine backends is flagged} by the deployment rule (0/9). Every provider stays below the mismatch count at which the $\mathrm{CP}_{99}$ binomial test would reject, so serving a quantized variant does not by itself change the audit verdict.
The error-rate gap introduced by quantization is small, so separating these backends requires a finer-grained comparison at the level of individual probes rather than aggregate error alone.
For this we apply a more sensitive test, McNemar's paired test, which compares per-probe agreement against the reference and separates three of the five GLM-4.7 providers (Venice fp4 $p = 6\times10^{-5}$, DeepInfra fp4 $p = 0.008$, AtlasCloud fp8 $p = 0.035$).
On DeepSeek-V3.2 it isolates the Novita provider ($p = 0.005$) despite its nominal fp8 label, revealing backend inconsistency that the declared quantization level alone does not expose.
Provider pinning in the main detection experiment (Section~\ref{sec:setup}) removes this source of backend variation from the 155-pair matrix.

\partitle{R4. Backend-provider robustness}
We audit Claude Sonnet~4.6 through five provider routes using a fixed KBF probe set and layout.
KBF accepts all five, with mismatch rates of 5.75\%--6.44\%.
IRIS rejects the Amazon Bedrock route (10/64 tells, $p=0.00124<0.01$), yielding one false rejection across the five routes.
These results support KBF's robustness to ordinary backend variation in this setting and show that such variation can trigger a false alarm in IRIS.

\begin{figure}[!htbp]
    \centering
    \includegraphics[width=\linewidth]{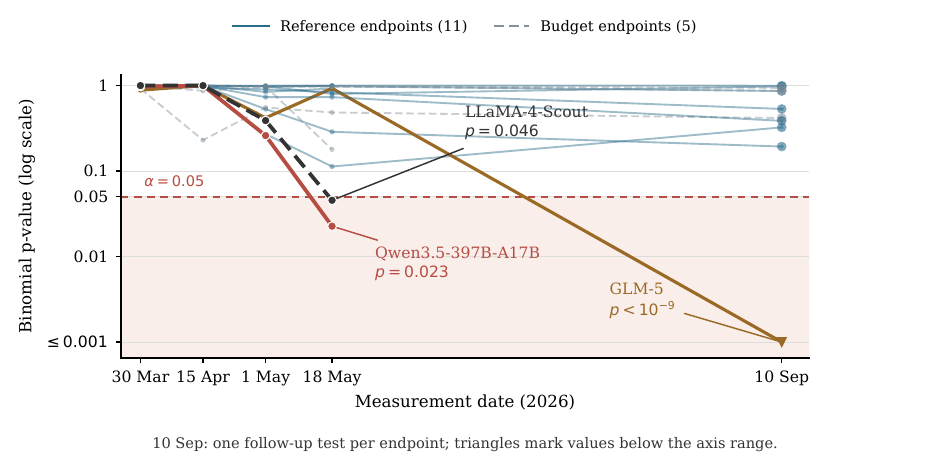}
    \caption{R5. Temporal stability from 15 March to 10 September 2026 against each endpoint's earliest available baseline. Solid and dashed curves denote reference and budget endpoints, respectively. The September follow-up samples 11 endpoints, with one test per endpoint. Highlighted curves mark rejected endpoints. Triangles place $p<0.001$ at the plotting floor. The dashed horizontal line marks $\alpha=0.05$. Baselines are omitted and missing tests break the curves.}
    \label{fig:rb_temporal}
\end{figure}

\partitle{R5. Temporal robustness}
Probe sets are generated from a reference snapshot, while providers may silently update their backends.
For all 16 models, we re-run each same-reference trial across five snapshots spanning 15 March to 18 May 2026 (64 days), applying the same $\mathrm{CP}_{99}$ binomial test as the main detection to each available snapshot against its earliest baseline (Figure~\ref{fig:rb_temporal}). Probes stay valid through 7 weeks, where \textbf{all 16 models remain stable}.
At the week-9 measurement, one of the eleven T1/T2 endpoints that can serve as a reference drifted, Qwen3.5-397B-A17B (self-error $1.7\% \to 7.8\%$, $p=0.023$), a shift we reproduced on the official API.
The rejection of LLaMA-4-Scout is borderline ($p=0.046$), consistent with its limited stability as a T3 model.
The observed increase in reference self-error motivates periodic retesting against the enrolled baseline and refreshing probes when reference behavior changes.
On 10 September 2026 (day 179), we retested a sample of 11 endpoints once against their original baselines. Ten remained consistent. GLM-5 showed increased self-error from $4.2\%$ to $16.5\%$, with $p<10^{-9}$. Gemini 2.5 Flash Lite had 209 scored responses after excluding one unparseable answer under the archived scoring convention.

\begin{figure}[!htbp]
    \centering
    \begin{minipage}[t]{0.48\linewidth}
        \centering
        \includegraphics[width=\linewidth]{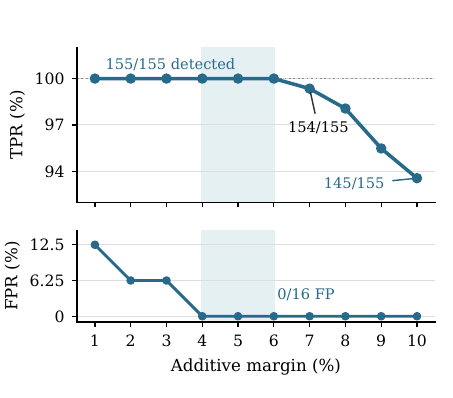}
        \caption{R6. Threshold sweep over 155 substitution pairs and 16 independent same-reference retests. Control rejections fall from 2/16 at 1\% to 1/16 at 2\%--3\% and 0/16 at 4\%--10\%.}
        \label{fig:rb_threshold}
    \end{minipage}\hfill
    \begin{minipage}[t]{0.48\linewidth}
        \centering
        \includegraphics[width=\linewidth]{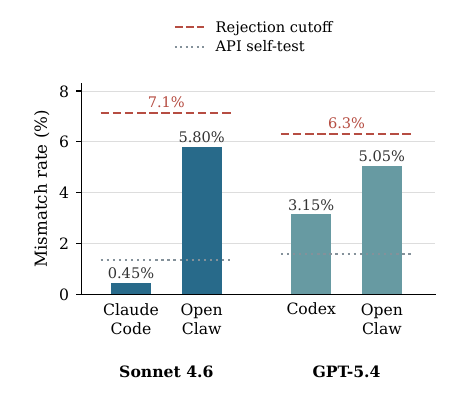}
        \caption{R7. Agent-interface audits. Bars show mismatch rates, dashed lines show rejection cutoffs, and dotted lines show API self-test rates. All four configurations pass.}
        \label{fig:rb_agentcli}
    \end{minipage}
\end{figure}

\partitle{R6. Threshold robustness}
The main evaluation uses the $\mathrm{CP}_{99}$ bound determined by each reference model's self-disagreement rate.
As a diagnostic check, we compare this rule against additive margins from 1\% to 10\% across all 155 economically relevant pairs (Figure~\ref{fig:rb_threshold}).

At margins of 1\%--6\%, all 155 pairs are detected (100\% TPR).
At a 7\% margin, detection remains at \textbf{99.4\% TPR} (154/155 TP); the single missed pair (Claude Sonnet~4.6 vs.\ Claude Opus~4.6) is a within-family pair that sits at the borderline and is correctly detected under the $\mathrm{CP}_{99}$ binomial test with the full probe set.
At 10\%, detection falls to 93.5\% TPR as the threshold absorbs the signal from moderately distinguishable pairs.
False positives are evaluated on 16 independent same-reference retests, separate from the self-tests used to set the baseline error rates.
There are 2/16 false positives at a 1\% margin, 1/16 at 2\%--3\%, and 0/16 at 4\%--10\%.
Thus, margins of 4\%--6\% retain all 155 detections with no observed false positives in these controls.

\partitle{R7. Agent-interface probing}
We finally test whether KBF stays stable under the system prompt of a real coding agent, so that a user can audit the API behind an agent simply by sending it the probes.
We evaluate two references (Claude Sonnet~4.6 and GPT-5.4) across three widely used agent CLIs (Claude Code, OpenClaw, and Codex).
On OpenClaw we connect both models for testing, while for the other two official CLIs we query the official APIs directly.
In each, the agent reads the probe document and answers inline within its own session, under the agent's full system prompt (the OpenClaw harness alone adds roughly 30~KB) and without temperature control; we then parse the replies and apply the standard test.
\textbf{All four configurations pass}: under the $\mathrm{CP}_{99}$ binomial test, every agent-side result stays consistent with the model's own API baseline and produces no false positive (0/4 FP), as Figure~\ref{fig:rb_agentcli} shows.
The agent system prompt is the dominant variable: the Claude Code session with an explicit recall-only instruction lands at 0.45\% mismatch, essentially the API baseline, while the two OpenClaw sessions sit near 5\% yet remain well inside their rejection thresholds.
A user can therefore audit the model behind an agent by sending it the probes directly, without API access.

\partitle{R8. Extended KBF Stress Test}
We extend the stress test in Section~\ref{sec:config-robustness} to the full model pool.
We sample five agent configurations from PromptConfFactory~\cite{llmmap}, including role prompts, chain-of-thought variants, and RAG injections, and apply each to all 16 models for 80 same-reference trials.
All T1 and T2 models remain stable with 0 FP.
GLM-4.7-Flash (T3), unstable because of its high baseline self-error, accounts for both false positives (2/80 trials, 2.50\% FPR).
This gives configuration-specific evidence that KBF remains conservative under ordinary endpoint wrappers.

\section{Partial-Routing Calibration and Extended Results}
\label{app:adaptive-routing-details}

\subsection{Experimental Setup}
\label{app:routing-drift}

We evaluate the eight reference--substitute pairs in Figure~\ref{fig:partial_routing_power} using the scoring procedure in Section~\ref{sec:kbf-routing-overview}.
For each reference, all comparisons use a fixed common subset of enrolled probes selected by reference agreement and response availability.
The replay preserves request groups of up to ten probes from one domain, reconstructed from archived domain labels and probe order under a consistent alignment assumption.

We reuse reference responses and substitute fingerprints from our earlier experiments.
For each request in each online round, an independent Bernoulli draw with probability $\pi$ selects the archived substitute response for the entire batch.
Otherwise, the replay uses the honest-reference response recorded for that round.
The reference anchor and candidate fingerprints remain fixed across replays.
We run 10,000 replays per pair and round count at each routing fraction, using a 1\% grid for detection and $\pi\in\{0,0.05,\ldots,1\}$ for estimation.
All replays enter the estimation analysis, regardless of whether the audit detects substitution.
The simulations condition on archived responses and involve no fresh API queries or reference recalibration.

\subsection{Intervals and Candidate Ranges}
\label{app:routing-intervals}

\partitle{Known substitute}
Let $\widetilde\pi_S=\bar d/\bar g$ be the unclipped ratio estimate from Section~\ref{sec:kbf-adaptive-routing}.
For $B>1$ and $\bar g>0$, the delta method gives the standard error
\[
    \widehat{\mathrm{SE}}_S
    =\sqrt{\frac{\sum_{b=1}^{B}(d_b-\widetilde\pi_S g_b)^2}
                       {B(B-1)\bar g^2}},
\]
where each request group is one variance unit.
For one round, $d_b=Y_{b1}-A_b$.
For two rounds, we average the mismatch fractions within each batch before computing $d_b$.

The nominal 90\% delta-method interval is
\[
    \mathcal J_S=\left[
        \widetilde\pi_S-1.645\,\widehat{\mathrm{SE}}_S,\;
        \widetilde\pi_S+1.645\,\widehat{\mathrm{SE}}_S
    \right]\cap[0,1],
\]
with empty intersections retained.
This approximation requires independent batches, stable reference means, and $\bar g$ sufficiently separated from zero.
The calculation conditions on the stored substitute fingerprint.

\partitle{Unknown substitute}
The candidate pool is $\mathcal C=\{$GLM-5, Qwen3.5-397B-A17B, Kimi-K2-0905, DeepSeek-V3.2$\}$.
We report the range of candidate point estimates defined in Section~\ref{sec:kbf-adaptive-routing}, provided all candidates have positive mean mismatch gaps and the pool contains the true substitute.
Budget-substitute evaluation reports known-model estimates.

\subsection{Extended Results}

\partitle{Partial-routing detection}
For one round, we use $T_1=\sum_{b=1}^{B}\operatorname{sgn}(Y_{b1}-A_b)$ and apply a one-sided paired sign test over non-tied batches.
Two rounds use the joint test in Section~\ref{sec:kbf-routing-overview}.
Figure~\ref{fig:partial_routing_power_one_round} reports one-round detection for the same pairs as the two-round results in Figure~\ref{fig:partial_routing_power}.
We define $\mathrm{MDR}_{95}$ as the first grid fraction reaching 95\% TPR under the paired test at $\alpha=0.05$.
For Opus $\rightarrow$ Kimi-K2, Sonnet $\rightarrow$ GLM-5, Gemini $\rightarrow$ Kimi-K2, and GPT-5.4 $\rightarrow$ Kimi-K2, one-round $\mathrm{MDR}_{95}$ values are 34\%, 75\%, 35\%, and 42\%, respectively.
Two rounds reduce them to 23\%, 55\%, 22\%, and 34\%.
The four budget substitutions reach 95\% TPR at 22--50\% routing with one round and 15--34\% with two.
A second round improves detection at the cost of additional audit queries.

\begin{figure}[!htbp]
    \centering
    \includegraphics[width=0.76\textwidth]{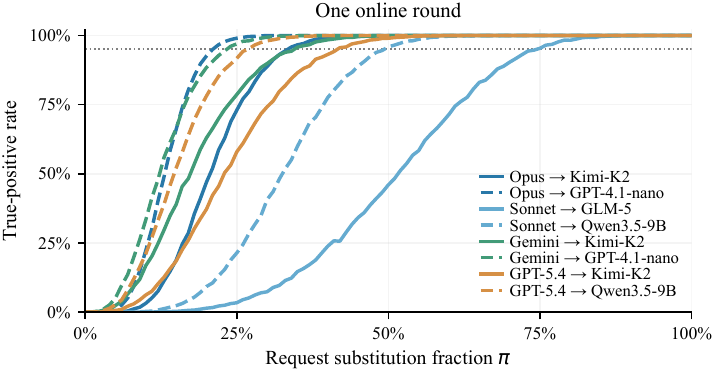}
    \caption{Paired batch detection with one online round. Solid curves denote same-class substitutes and dashed curves budget substitutes. Each point uses 10,000 routing replays at $\alpha=0.05$, conditional on archived responses. The dotted line marks 95\% TPR. Figure~\ref{fig:partial_routing_power} gives the two-round results for the same eight pairs.}
    \Description{One-round true-positive rates for eight whole-request substitution pairs as the routing fraction increases.}
    \label{fig:partial_routing_power_one_round}
\end{figure}

\begin{figure}[!htbp]
    \centering
    \includegraphics[width=\textwidth]{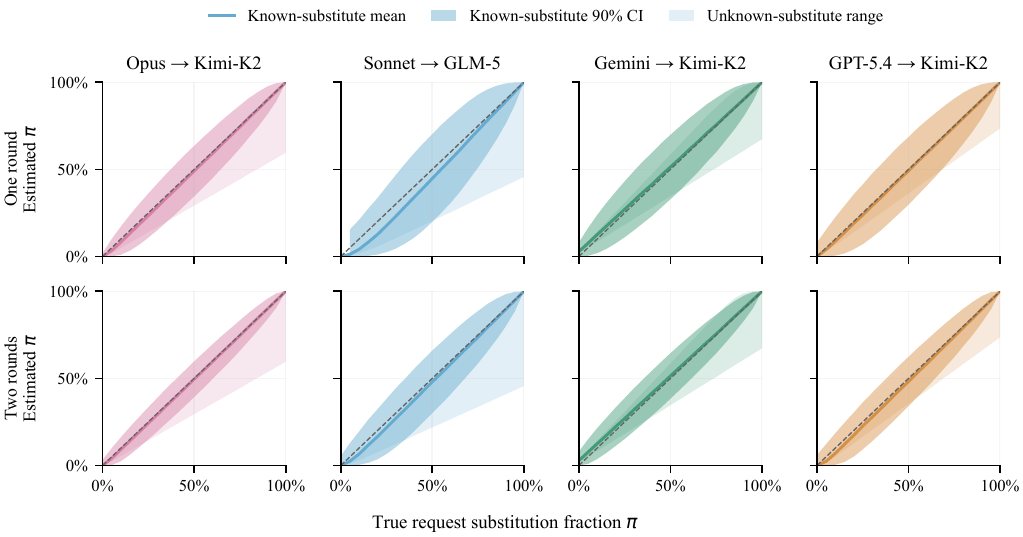}
    \caption{Routing-fraction estimation with unknown substitutes, with one round (top) and two rounds (bottom). Light bands connect mean minima and maxima of candidate point estimates. For comparison, solid lines show mean estimates for known substitutes, and dark bands connect mean endpoints of their nonempty nominal 90\% delta-method intervals. Dashed lines show $y=x$.}
    \Description{Four columns show Opus, Sonnet, Gemini, and GPT-5.4 with their same-class substitutes. The top and bottom rows show one and two online rounds, with known-substitute intervals and candidate-pool point-estimate ranges.}
    \label{fig:pi_hat}
\end{figure}

\partitle{Unknown-substitute estimation}
Figure~\ref{fig:pi_hat} compares one- and two-round routing-fraction estimates for four same-class pairs, using known-substitute estimates as a reference.
The spread across candidate estimates reflects uncertainty about substitute identity.
A second round improves substitution detection and narrows the nominal intervals for known substitutes.

\section{Real-World Relay Audits: Extended Results}
\label{app:shadow-details}

This appendix expands the field-audit results of Section~\ref{sec:shadow-api} to all eleven audited platforms.

\partitle{Per-model overview across all eleven platforms}
Figure~\ref{fig:shadow_per_model} summarizes the per-model consistency split across all eleven audited platforms, where each bar covers every platform that exposes the model.
The full survey shows the same concentration pattern as the representative subset.
Claude Opus~4.6 is flagged on 7 of 10 platforms that serve it and Claude Sonnet~4.6 on 3 of 11, so the two most expensive proprietary models account for \textbf{10 of the 12} flagged endpoints across the eleven platforms.
GLM-4.7 is flagged on 2 of 5 platforms, while GPT-5.4, DeepSeek-V3.2, and Gemini~3 Flash are consistent on every platform that exposes them.
The aggregate flag rate across all eleven platforms is \textbf{12/42}, measuring statistical inconsistency with the probe-response distributions of the pinned references in this sample.

\begin{figure}[!htbp]
    \centering
    \includegraphics[width=0.6\textwidth]{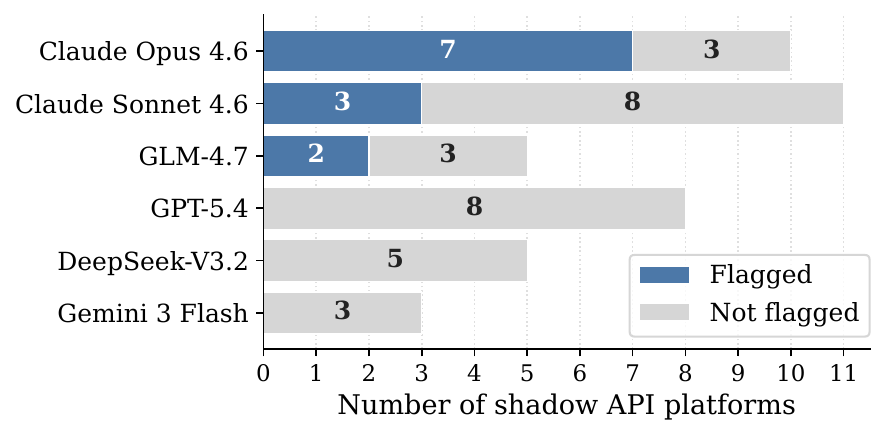}
    \caption{Per-model consistency split across the eleven-platform shadow API field audit. Each bar covers all platforms that expose the model.}
    \Description{Horizontal stacked bar chart with six rows, one per audited model. Claude Opus is flagged on seven platforms, Claude Sonnet on three, GLM-4.7 on two, and GPT-5.4, DeepSeek-V3.2, Gemini 3 Flash are entirely consistent.}
    \label{fig:shadow_per_model}
\end{figure}

To assess repeatability, the high-mismatch Claude endpoints on Platform~1 and Platform~5 are re-measured across three independent rounds, while the rest are single-round.

\partitle{Platform Mapping}
\label{app:platform-mapping}
Table~\ref{tab:platform-mapping} maps the platform identifiers used in Section~\ref{sec:shadow-api} to their salted SHA-256 digests.

\begin{table}[h]
    \centering
    \small
    \setlength{\tabcolsep}{8pt}
    \caption{Platform identifiers and their salted SHA-256 digests.}
    \label{tab:platform-mapping}
    \begin{tabular}{@{}lll@{}}
        \toprule
        P1\ \ \texttt{SH-87b19c56} & P2\ \ \texttt{SH-3c82a49c} & P3\ \ \texttt{SH-4f754cc3} \\
        P4\ \ \texttt{SH-75f73c6e} & P5\ \ \texttt{SH-0989276f} & P6\ \ \texttt{SH-031d44cf} \\
        P7\ \ \texttt{SH-3607d87e} & P8\ \ \texttt{SH-1a4df8c4} & P9\ \ \texttt{SH-feecd406} \\
        P10\ \ \texttt{SH-7e45d454} & P11\ \ \texttt{SH-bd05fa88} & \\
        \bottomrule
    \end{tabular}
\end{table}

\section{Baseline Configuration and Related Work}
\label{app:baselines-related}

\subsection{Baseline Configuration}
\label{app:baseline-details}

LLMmap, MET, ZeroPrint, and IRIS cover the 16-model pool, while GateScope covers a 12-model subset.
We use the released implementations with the adaptations described below to decide whether a suspect endpoint is consistent with its claimed reference.
Provider pins remain fixed within each collection.
The later IRIS and GateScope runs use Novita for Kimi-K2 and LLaMA-4-Scout for these later collections.

\partitle{LLMmap}
LLMmap is an enrollment-and-retrieval fingerprint. We extend the pretrained open-set library with 16 new model templates, and at test time the unknown endpoint answers 8 queries in the bare-query setting at $\mathrm{temp}{=}0$. LLMmap converts the responses into a feature vector and ranks all enrolled templates by Euclidean distance.\footnote{We follow the public implementation and use Euclidean distance for the ranking computation.} An endpoint is consistent with its claimed reference when that reference is its Top-1 nearest template, and any other Top-1 match flags it as a different model.

\partitle{MET}
MET is a two-sample distributional test. We use 25 multilingual Wikipedia continuation prompts drawn from a 500-record candidate pool, and for each model collect two independent batches of 250 completions (10 per prompt). The test statistic is MMD with a Hamming kernel, and $p$-values are estimated from 1000 permutations. An endpoint is consistent with its claimed reference when the two-sample test does not reject the same-model null at $\alpha{=}0.05$, and a rejection flags it as a different model.

\partitle{ZeroPrint}
ZeroPrint uses structured perturbations to estimate local response sensitivity, building a Jacobian-style fingerprint compared by Pearson similarity. Following the original method, we issue each prompt and its perturbed variants 20 times, for 200 requests per model. Our adaptation is the relay-auditing decision. The predefined similarity threshold from the original evaluation is fixed on base-and-derivative model pairs and does not suit this setting, so we instead fingerprint each endpoint in two independent collections and build a reference library from the first. We then pool all query--reference similarity scores and set a single global decision threshold at the Youden-optimal operating point, matching the original protocol of choosing one threshold over the full score set. An endpoint is consistent with its claimed reference when its similarity to that reference's library fingerprint meets the global threshold, and a similarity below the threshold flags it as a different model.

\partitle{IRIS}
We use the publicly released IRIS code for reference enrollment and endpoint auditing.
At $\mathrm{temp}{=}1$, we collect 100 enrollment responses per model for each of six random-generation prompts, 120 independent calibration responses, and 64 audit responses cycling through the six prompts.
We train a random forest on content-only features pooled across the prompts and calibrate its threshold using independent honest responses.
Under the released decision rule, a response is a tell when its negative log posterior for the claimed model exceeds the 95th percentile of the calibration scores.
The audit rejects the same-model null when an upper-tail binomial test of the tell count against a nominal honest rate of 0.05 gives $p<0.01$.
Robustness tests keep enrollment and calibration fixed and collect 64 responses under each deployment configuration.

\partitle{GateScope}
GateScope uses 55 capability probes with a JSON system prompt at $\mathrm{temp}{=}0.7$ and one-vs-rest XGBoost classifiers.
We retain vendor-default reasoning and use six repetitions per probe, split into three for training, one for probe selection, and two for calibration.
The evaluated subset excludes Claude Opus~4.6, GPT-5.4, Qwen3.5-397B-A17B, and GLM-4.7-Flash.
We implement selection of 12 probes per claimed model and add a decision rule to the original attribution scores.
For each claimed model, 24 held-out responses estimate the honest attribution rate, and 60 independent audit responses are tested against it using a lower-tail binomial test at $\alpha{=}0.01$.
Enrollment and audit collection for the 12 evaluated models cost \$29.78.

\subsection{Black-Box Auditors}
\label{app:auditors}

Black-box auditors differ in their probe signals, decision rules, and reference requirements. Table~\ref{tab:auditors} summarizes these properties. Appendix~\ref{app:baseline-details} describes the baselines evaluated in this paper.

\begin{table}[!htb]
    \centering
    \small
    \renewcommand{\arraystretch}{1.1}
    \setlength{\tabcolsep}{2pt}
    \caption{Black-box auditors. \textbf{Probe}: univ = one fixed set for all models, per-ref = generated for the claimed reference, natural = ordinary user prompts. \textbf{Decision}: test = hypothesis test, classifier = learned classifier or nearest neighbor, threshold = similarity or distance cutoff. \textbf{FPR}: the rule bounds the false-positive rate. \textbf{Routing}: detects partial substitution.}
    \label{tab:auditors}
    \begin{tabularx}{\linewidth}{@{}>{\raggedright\arraybackslash}p{3.1cm} >{\raggedright\arraybackslash}X >{\raggedright\arraybackslash}p{0.95cm} >{\raggedright\arraybackslash}p{1.35cm} c c >{\raggedright\arraybackslash}p{2.15cm} c@{}}
        \toprule
        \textbf{Method} & \textbf{Signal} & \textbf{Probe} & \textbf{Decision} & \textbf{FPR} & \textbf{Temp} & \textbf{Reference} & \textbf{Routing} \\
        \midrule
        \textbf{\name (ours)} & knowledge boundary & per-ref & test & Y & any & official API & Y \\
        LLMmap\newline\cite{llmmap} & prompt bank & univ & classifier & N & any & enrolled library & N \\
        MET\newline\cite{met} & completion distribution & natural & test & Y & $>0$ & reference samples & N \\
        ZeroPrint\newline\cite{zeroprint} & input perturbation & univ & threshold & N & any & reference fingerprint & N \\
        FDLLM\newline\cite{fu2025fdllm} & generated text & univ & classifier & N & any & labeled corpus & N \\
        GateScope\newline\cite{lin2026behavioral} & capability probes & univ & classifier & N & 0.7 & official API corpus & N \\
        RUT\newline\cite{zhu2026auditing} & completion distribution & natural & test & Y & 0.5 & local open weights & Y \\
        AgentProv\newline\cite{wang2026agentprov} & tool choice & univ & test & Y & 0.7 & reference instance & Y \\
        IRIS\newline\cite{zhang2026model} & random generation & univ & test & Y & $>0$ & reference + candidates & Y \\
        One Token\newline\cite{bruckner2026one} & random generation & univ & threshold & N & 1 & registered deployment & N \\
        FLIPS\newline\cite{richardeau2026flips} & random generation & univ & classifier & N & $>0$ & enrolled instances & N \\
        Ventor-QTest\newline\cite{wu2026ventor} & random generation & univ & test & Y & 1 & official API + logprobs & N \\
        IKP\newline\cite{li2026incompressible} & factual recall & univ & threshold & N & 0 & open-weight calibration & N \\
        \bottomrule
    \end{tabularx}
\end{table}

\partitle{Prompt banks and classifiers}
LLMmap sends eight fixed queries, encodes the answers with a trained trace encoder, and returns the nearest enrolled template~\cite{llmmap}. FDLLM trains a detector on generations from a closed model set~\cite{fu2025fdllm}, and GateScope classifies JSON answers to competition and factual questions with one XGBoost model per endpoint, then reports the share of gateway responses attributed to the advertised model~\cite{lin2026behavioral}. These methods identify enrolled models quickly, but the verdict is a label or a distance with no bound on how often an honest endpoint is flagged, and the fixed query set can be recognized and forwarded to the genuine backend. GhostPrint shows that a cheaper model can be fine-tuned to pass such signatures~\cite{zhang2026your}.

\partitle{Distribution tests}
MET samples continuations of Wikipedia prefixes from both endpoints and applies a two-sample MMD test~\cite{met}. RUT sends each prompt once, scores the response by its log-rank under a locally hosted reference, and tests the ranks for uniformity, which needs open weights on the reference side~\cite{zhu2026auditing}. AgentProv applies the same two-sample idea to which of several equivalent tools the model calls~\cite{wang2026agentprov}. All three use calibrated tests, but MET remains sensitive to deployment changes. Section~\ref{sec:config-robustness} measures this for MET.

\partitle{Random-generation bias}
Asked for a random integer or bit string, a model produces a skewed distribution that identifies its backend. IRIS turns such strings into features, scores each response by how atypical it is for the claimed model, and adds a tell-count test for partial routing and a pilot that fixes the budget in advance~\cite{zhang2026model}. Bruckner compares single-word answer distributions by Jensen--Shannon distance~\cite{bruckner2026one}, FLIPS classifies NIST randomness statistics of generated bit sequences~\cite{richardeau2026flips}, and Ventor-QTest compares single-token draws in frozen contexts with a coarsened KL statistic~\cite{wu2026ventor}. These probes require no per-model mining. The signal, however, lives in the tail of the sampling distribution, so it needs a known nonzero temperature and shifts with quantization and kernels. IRIS separates fourteen of fifteen same-model provider pairs, Bruckner finds ten of thirty-four such pairs farther apart than different models, and Wimbauer et al.\ show that the inference engine and GPU leave readable traces in text~\cite{wimbauer2026fingerprinting}. A random-generation fingerprint therefore measures the serving configuration together with the weights. \name compares greedy answers, which these changes flip far less often.

\partitle{Perturbation and change detection}
ZeroPrint fits a local Jacobian from paraphrased inputs to response embeddings and compares endpoints by correlation, which depends on the embedding model and on a threshold that must hold across configurations~\cite{zeroprint}. B3IT, Log Probability Tracking, and the Stability Monitor compare an endpoint with its own past, through border inputs at zero temperature, first-token log probabilities, or embedding distances aggregated over time~\cite{chauvin2026token,chauvin2026log,leshin2026behavioral}. These methods use historical behavior as the baseline for detecting endpoint changes. Frequent changes in relay backends and serving configurations make it difficult to collect a stable history of the claimed model at a relay endpoint. These methods are therefore poorly suited to black-box verification in our relay setting, and we exclude them from the empirical comparison.

\partitle{Knowledge-based signals}
Knowledge-boundary work asks which facts a model holds and how it should answer beyond them~\cite{li2025knowledge,deng2025unveiling,wang2026knowledge}. Concurrent work on Incompressible Knowledge Probes estimates model size from recall across rarity tiers and uses shared wrong answers as a lineage signal~\cite{li2026incompressible}. \name independently develops a knowledge-boundary fingerprint for relay auditing, with per-reference probes, greedy matching, and a binomial test whose null is calibrated on the reference itself.

\subsection{Additional Related Work}
\label{app:related}

The methods below either need more than text from the audited endpoint or need another party to act first.

\partitle{Adversarial suffixes and planted fingerprints}
TRAP optimizes a suffix to induce a chosen completion~\cite{gubri2024trap}, RoFL optimizes prompts for model identification~\cite{rofl}, and LLMPrint injects instructions that force model-specific token preferences~\cite{llmprint}. The probes are built with access to the source model and look nothing like ordinary traffic, so a relay can refuse them or forward them to the genuine backend~\cite{zeroprint}. Planted fingerprints embed ownership signals through training or fine-tuning~\cite{xu2024instructional,adi2018turning,gloaguen2025robust,xu2025evertracer,russinovich2024hey}. They settle ownership disputes over stolen weights, which a relay user never holds.

\partitle{Internal and probability channels}
Attention and gradient fingerprints need white-box access~\cite{zhang2026attndiff,wu2025gradient}. Logit-space signatures work through an API that returns probabilities~\cite{yang2024fingerprint,finlayson2025every}. Relays rarely pass probabilities through, and closed providers return at most a truncated top-$k$.

\partitle{Watermarks and verifiable inference}
Sampling watermarks introduce detectable statistical patterns at the token or semantic level~\cite{kirchenbauer2023watermark,dathathri2024scalable,hou2024semstamp,huo2025pmark}. They answer whether text came from a watermarked model, and they help a relay user only if the official provider watermarks every response and exposes the detector. Zero-knowledge proofs, trusted hardware, and selective verification prove that an inference ran on the claimed model~\cite{sun2024zkllm,maheri2025telesparse,tramer2018slalom,sabt2015trusted,guo2026immaculate}. These give stronger guarantees than any behavioral test and start working only after the provider deploys them. \name covers the intermediary that has deployed nothing.

\Needspace{6\baselineskip}
\section{Ablation Study}
\label{app:signal-ablation}

\newcommand{\eThirteenLegend}[1]{%
    {\small\textcolor[HTML]{A0AD55}{\rule{6pt}{6pt}}\enspace Full method
    \qquad\textcolor[HTML]{E7A34B}{\rule{6pt}{6pt}}\enspace #1}\par\smallskip
}

\begin{mdframed}[linewidth=0.7pt,linecolor=black!65,backgroundcolor=black!3,
    innerleftmargin=9pt,innerrightmargin=9pt,innertopmargin=7pt,innerbottommargin=7pt,
    skipabove=7pt,skipbelow=9pt,nobreak=true]
\partitle{Key takeaway}
With the same number of generation rounds, specialist facts yield more retained probes and larger agreement gaps for most tested model pairs.
Stability filtering lowers reference noise, while contrast filtering improves model discrimination, supporting detection at a fixed audit budget.
Replacing Qwen3.5-9B with Phi-4 preserves detection performance across all tested model pairs.
\end{mdframed}

\subsection{Components and Evaluation}
\label{app:signal-ablation-setup}
\label{app:signal-ablation-components}

\partitle{Specialist candidate generation}
The reference proposes numerical facts from specialist topics within each domain.
This component controls which facts enter the candidate pool.
Its ablation replaces the specialist prompts and themes with common factual topics in the same domains, while retaining the question format and numerical comparison rules.

\partitle{Stability filtering}
The reference answers each candidate once at temperature~0 and twice at temperature~0.7.
The filter retains candidates whose three responses satisfy the enrollment consensus rule.
Its ablation removes this agreement requirement while retaining contrast filtering.

\partitle{Contrast filtering}
Qwen3.5-9B answers the same candidates.
The filter retains a candidate when the contrast answer is parseable and differs from the temperature-0 reference answer under the domain comparison rule.
Its ablation omits this test while retaining stability filtering.

The full method combines specialist candidate generation with both filters.
Table~\ref{tab:e13-single-components} specifies the three ablations.
The three component ablations hold difficulty at the initial level and register the temperature-0 reference answer, so a filter ablation changes probe selection without changing the answer used for scoring.

\begin{table}[!htbp]
    \centering
    \small
    \setlength{\tabcolsep}{7pt}
    \caption{Each ablation changes one component of the full method. Common-fact generation replaces specialist topic selection and retains both filters.}
    \label{tab:e13-single-components}
    \begin{tabular}{@{}llcc@{}}
        \toprule
        Configuration & Candidate topics & Stability filter & Contrast filter \\
        \midrule
        Full method & Specialist & On & On \\
        Common facts & Common & On & On \\
        Without stability filtering & Specialist & Off & On \\
        Without contrast filtering & Specialist & On & Off \\
        \bottomrule
    \end{tabular}
\end{table}

\partitle{Data and metrics}
We evaluate Claude Sonnet~4.6, GPT-5.4, DeepSeek-V3.2, and GPT-4.1-mini as references, with the other three models serving as targets for each reference.
Calibration, reference self-evaluation, and target evaluation use separate response batches.
Agreement measures reproducibility of the registered answers rather than factual correctness, with invalid responses counted as mismatches.
Model discrimination is the reference agreement rate minus the target agreement rate for each model pair.

\partitle{Generation budget and evaluation coverage}
The generation comparison uses the first four rounds in each of 15 domains, totaling 60 rounds per reference and arm.
Round selection follows generation order without conditioning on yield or evaluation outcomes.
Independent calibration, reference self-evaluation, and all three target evaluations cover every retained common-fact probe in the 60-round pool, while specialist coverage remains partial for three references.
Figure~\ref{fig:e13-ablate-generation} reports yields and pairwise agreement gaps for this 60-round comparison.

\partitle{Screening audit budget}
Screening variants select probes from the same evaluated fixed-difficulty candidate set for each reference.
We evaluate 200-probe audits across 100 selection seeds, using separate 200-probe calibration batches, a model-specific $\mathrm{CP}_{99}$ null bound, and the one-sided binomial test at $\alpha=0.05$.
Each directed reference--target pair therefore has 100 audit decisions.
We report each directed pair separately and evaluate self-controls separately for each reference.
The seeds select probe subsets from stored responses.
The screening ablations report target rejection percentages for individual directed model pairs at 200 probes, restricted to the evaluated candidate set.
The contrast-model comparison in Section~\ref{app:signal-ablation-contrast-model} uses its separately specified protocol.

\Needspace{6\baselineskip}
\subsection{Ablating Specialist Candidate Generation}
\label{app:signal-ablation-generation}
\label{app:signal-ablation-common-facts}

\partitle{Comparison}
We replace specialist topic selection with common factual topics while retaining both filters.
Both arms use the same 15 domains, numerical question templates, matching tolerances, and 60 generation rounds.
For example, the astronomy control requests orbital and rotation periods of planets and major moons rather than measurements of obscure objects.
Yield counts all retained probes, while agreement gaps use those with complete independent evaluations.
Specialist evaluation covers 217/351, 338/338, 188/307, and 125/191 retained probes for Sonnet~4.6, GPT-5.4, DeepSeek-V3.2, and GPT-4.1-mini, respectively, while common-fact evaluation covers every retained probe.

\partitle{Effect on probe yield}
At the same 60 generation rounds, specialist generation meets our minimum requirement of 100 retained probes for every reference (Figure~\ref{fig:e13-ablate-generation-a}).
Common-fact generation reaches this count for three references, while DeepSeek-V3.2 yields 89 probes.
For Sonnet~4.6, common facts yield 137 retained probes versus 351 with specialist topics.
For GPT-5.4, the corresponding counts are 135 and 338.
Across the four references, specialist generation supplies 1.71--3.45 times as many probes passing both filters.
Across references, 69.33--74.26\% of valid common-fact candidates pass stability filtering but share their registered answer with Qwen3.5-9B, so contrast filtering removes them.
These counts compare generation rounds, not token or monetary cost.

\begin{figure}[!htbp]
    \centering
    \eThirteenLegend{Common facts, both filters retained}
    \captionsetup[subfigure]{position=top,font=small,skip=3pt}
    \begin{subfigure}[t]{.42\linewidth}
        \centering
        \caption{Yield at 60 rounds}
        \label{fig:e13-ablate-generation-a}
        \includegraphics[height=6.8cm]{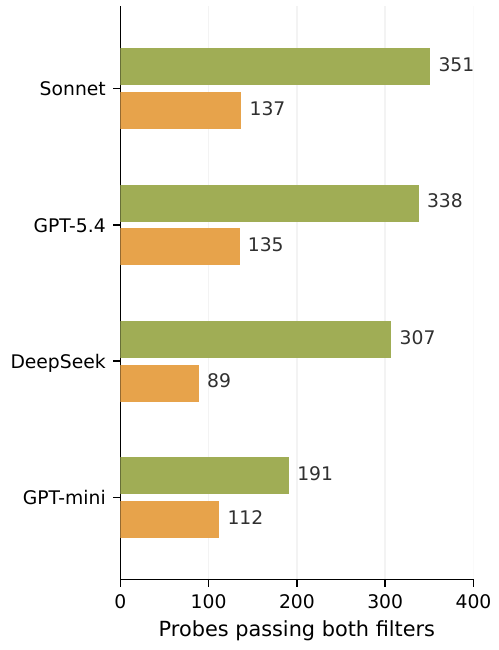}
    \end{subfigure}\hfill
    \begin{subfigure}[t]{.56\linewidth}
        \centering
        \caption{Pairwise agreement gap}
        \label{fig:e13-ablate-generation-b}
        \includegraphics[height=6.8cm]{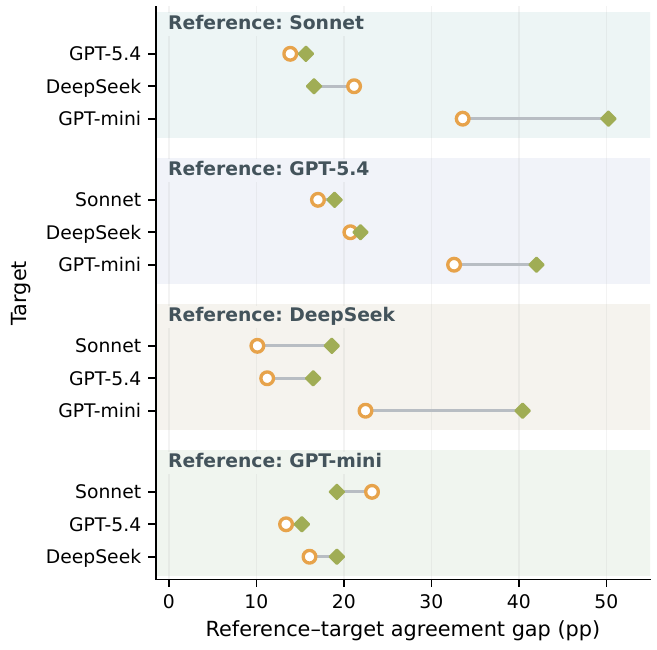}
    \end{subfigure}
    \caption{Specialist versus common-fact generation, with both filters retained. (a) Retained counts from the first four rounds per domain. (b) Reference self-agreement minus target agreement for each directed pair, in percentage points. Shaded groups identify references and rows identify targets. Green diamonds denote specialist probes and yellow circles common facts.}
    \label{fig:e13-ablate-generation}
\end{figure}

\partitle{Effect on model discrimination}
Common-fact probes also provide smaller agreement gaps for several model pairs (Figure~\ref{fig:e13-ablate-generation-b}).
With Sonnet~4.6 as the reference and GPT-4.1-mini as the target, the gap is 33.58 percentage points for common facts versus 50.23 for specialist topics.
With GPT-5.4 as the reference and the same target, the gap is 32.59 versus 42.01 percentage points.
Specialist probes have larger gaps in 10 of the 12 directed pairs, so the improvement depends on the target.
The consistent benefit of specialist generation is its larger retained pool under the same number of generation rounds.
Data were collected at different times, and specialist evaluation remains partial for three references, so the comparison does not isolate topic selection from temporal or serving variation.

\Needspace{6\baselineskip}
\subsection{Ablating Stability Filtering}
\label{app:signal-ablation-stability}

\partitle{Comparison}
We remove only the requirement that the reference reproduce its answer across the three construction queries.
Specialist candidate generation, contrast filtering, and the registered temperature-0 answer remain unchanged.
Both configurations use the same evaluated candidate frame and the same 200-probe audit budget.

\partitle{From reference stability to the audit decision}
The audit rejects a target when its mismatch count exceeds what the calibrated reference noise can explain.
For a fixed probe budget and significance level, a larger null bound $p_0$ requires more mismatches to reject the target.
Stability filtering therefore affects the final decision through reference calibration, as well as through the reference--target agreement gap.
Removing the filter raises the calibrated null bound for each reference at 200 probes (Figure~\ref{fig:e13-ablate-stability-a}).
The audit consequently tolerates more target mismatches before reporting \textsc{Different}.

\begin{figure}[!htbp]
    \centering
    \eThirteenLegend{Without stability filtering}
    \captionsetup[subfigure]{position=top,font=small,skip=3pt}
    \begin{subfigure}[t]{.42\linewidth}
        \centering
        \caption{Reference noise bound}
        \label{fig:e13-ablate-stability-a}
        \includegraphics[height=6.8cm]{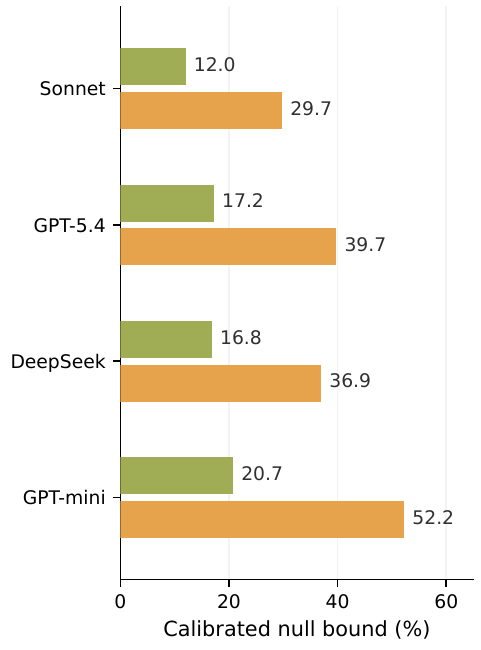}
    \end{subfigure}\hfill
    \begin{subfigure}[t]{.56\linewidth}
        \centering
        \caption{Pairwise detection (200 probes)}
        \label{fig:e13-ablate-stability-b}
        \includegraphics[height=6.8cm]{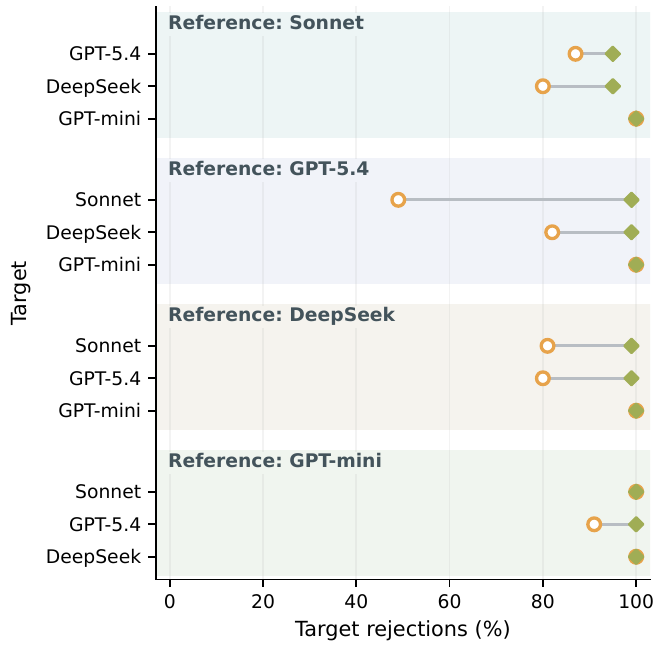}
    \end{subfigure}
    \caption{Effect of stability filtering at 200 probes. Green denotes the full method and yellow removal of the filter. (a) Mean calibrated $\mathrm{CP}_{99}$ reference mismatch bound across 100 selection seeds. (b) Target rejections per directed pair over those 100 seeds at $\alpha=0.05$. Shaded groups identify references and rows identify targets. Seeds reuse stored responses, and each reference has 0/100 self-control rejections under either configuration.}
    \label{fig:e13-ablate-stability}
\end{figure}

\partitle{Effect on discrimination and detection}
The reference--target agreement gap alone does not determine the decision, which also depends on the calibrated null bound and the probe budget.
With GPT-5.4 as the reference and Sonnet~4.6 as the target, removing stability filtering reduces rejections from 99/100 to 49/100 selection seeds at 200 probes (Figure~\ref{fig:e13-ablate-stability-b}).
The reverse direction, with Sonnet~4.6 as the reference, falls from 95/100 to 87/100.
Each reference has 0/100 self-control rejections under either configuration.
Stability filtering thus supports the final judgment by reducing reference noise and improving detection at a fixed audit budget.

\Needspace{6\baselineskip}
\subsection{Ablating Contrast Filtering}
\label{app:signal-ablation-contrast}

\partitle{Comparison}
We remove only the requirement that Qwen3.5-9B give a different answer.
Specialist candidate generation, stability filtering, and the registered answer remain unchanged.
The comparison again uses the same candidate frame and a 200-probe audit budget.

\partitle{Effect on model discrimination}
Without contrast filtering, the reference--target agreement gap decreases for every directed pair.
The retained answers are more widely shared across models.
Figure~\ref{fig:e13-ablate-contrast-a} summarizes target mismatch separately for each reference as a probe diagnostic.
The detection results in panel~(b) retain every target separately.

\begin{figure}[!htbp]
    \centering
    \eThirteenLegend{Without contrast filtering}
    \captionsetup[subfigure]{position=top,font=small,skip=3pt}
    \begin{subfigure}[t]{.42\linewidth}
        \centering
        \caption{Target mismatch}
        \label{fig:e13-ablate-contrast-a}
        \includegraphics[height=6.8cm]{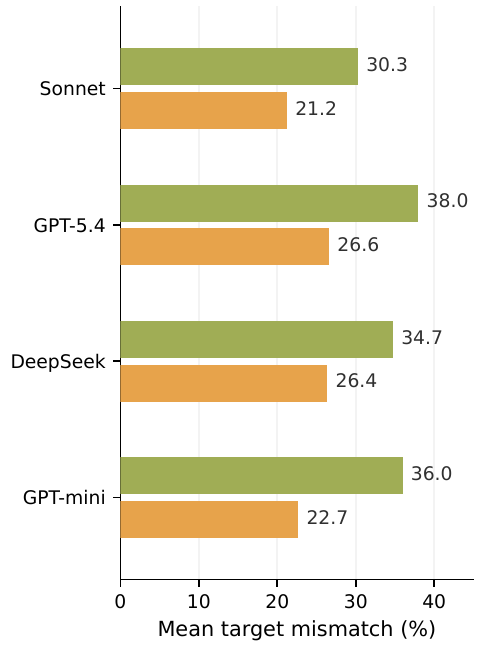}
    \end{subfigure}\hfill
    \begin{subfigure}[t]{.56\linewidth}
        \centering
        \caption{Pairwise detection (200 probes)}
        \label{fig:e13-ablate-contrast-b}
        \includegraphics[height=6.8cm]{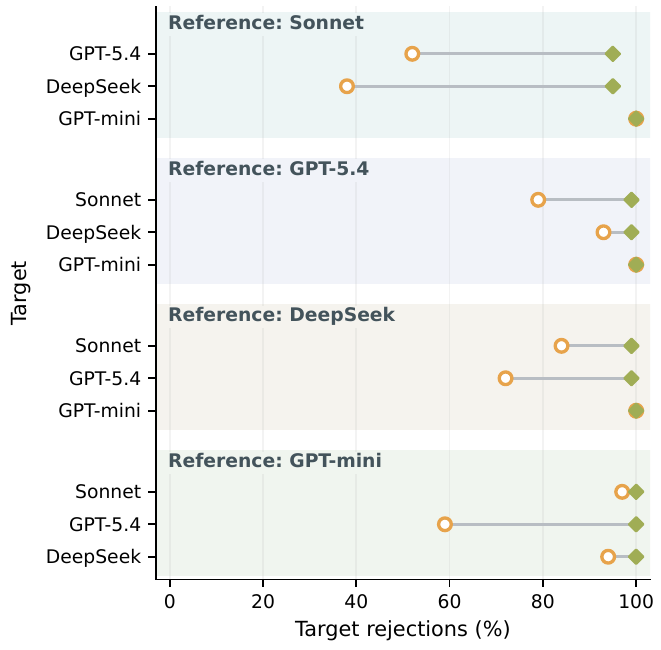}
    \end{subfigure}
    \caption{Effect of contrast filtering. Green denotes the full method and yellow removal of the filter. (a) Mean target mismatch over all retained probes in the evaluated frame. (b) Target rejections per directed pair over 100 selection seeds, using 200 audit probes, 200 independent calibration responses, and $\alpha=0.05$. Shaded groups identify references and rows identify targets. Seeds reuse stored responses, and each reference has 0/100 self-control rejections under either configuration.}
    \label{fig:e13-ablate-contrast}
\end{figure}

\partitle{Effect on detection}
For Sonnet~4.6 as the reference and DeepSeek-V3.2 as the target, rejections fall from 95/100 to 38/100 (Figure~\ref{fig:e13-ablate-contrast-b}).
For Sonnet~4.6 as the reference and GPT-5.4 as the target, rejections fall from 95/100 to 52/100.
Each reference has 0/100 self-control rejections under either configuration.
The small improvement in reference stability does not compensate for the loss of model discrimination.

\Needspace{6\baselineskip}
\subsection{Choice of Contrast Model}
\label{app:signal-ablation-contrast-model}

\partitle{Comparison and evaluation}
We replace Qwen3.5-9B with Phi-4 or Llama-3.1-8B-Instruct when screening the stored candidates, while retaining specialist candidate generation and stability filtering.
A fourth arm omits contrast filtering.
This control uses the same four references and their 1,600-question evaluation frames, with progressive candidate generation, consensus answers, and up to 400 retained probes per arm over ten selection seeds.
Each arm uses separate calibration responses to compute its own $\mathrm{CP}_{99}$ bound and applies the one-sided binomial test at $\alpha=0.05$, excluding unparseable responses from each batch.
These settings differ from the fixed-difficulty, 200-probe screening ablations above.
For each reference, we evaluate the other three references and independent Qwen3.5-9B target answers.
The selection seeds sample probe subsets from stored responses.

\partitle{Effect on detection}
Replacing Qwen3.5-9B with Phi-4 preserves detection for every tested directed pair, with both screening models yielding 10/10 rejections per pair.
Llama screening yields 0/10 rejections for Sonnet~4.6 $\rightarrow$ GPT-5.4 and 10/10 for every other pair.
Without contrast filtering, Sonnet~4.6 $\rightarrow$ GPT-5.4 and Sonnet~4.6 $\rightarrow$ DeepSeek-V3.2 yield 7/10 and 9/10 rejections, respectively, while all other pairs yield 10/10.
Every arm detects Qwen3.5-9B in 10/10 seeds for each reference, including when another model performs contrast screening.
Each reference has 0/10 self-control rejections under every arm.

\partitle{Effect on probe selection}
Different contrast models select different probe sets even when their detection results agree (Figure~\ref{fig:e13-contrast-overlap}).
The Qwen and Phi-4 sets have Jaccard similarity of 0.49--0.59, measured as the size of their intersection divided by the size of their union.
Llama retains 151--197 probes per reference, compared with 210--376 for Qwen and 263--373 for Phi-4, with Qwen--Llama similarity of 0.25--0.30.
All three contrast-screened sets fit below the cap.
The comparison therefore measures the combined effect of changes in probe membership, retained count, and calibration, rather than isolating contrast-model identity at a fixed probe count.

\begin{figure}[!htb]
    \centering
    \includegraphics[width=\linewidth]{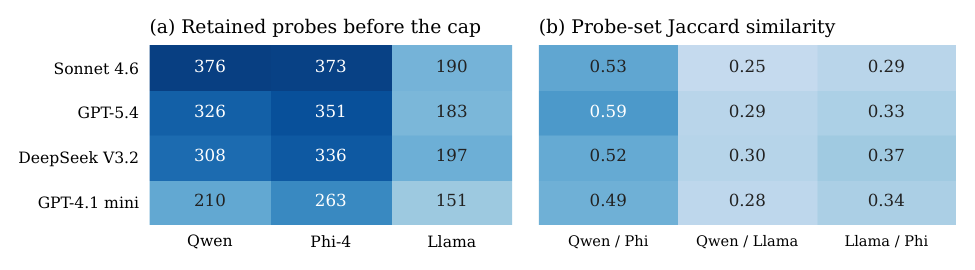}
    \caption{Effect of contrast-model choice on probe selection within the common evaluation frame. (a) Number of retained probes for each reference and contrast model. (b) Pairwise Jaccard similarity of the selected probe sets, where 1 denotes identical sets and 0 denotes disjoint sets. Rows identify the same references in both panels.}
    \label{fig:e13-contrast-overlap}
\end{figure}

\Needspace{24\baselineskip}
\Needspace{6\baselineskip}
\section{Limitations and Future Work}
\label{app:limitations}
\label{sec:operational-limitations}

\partitle{Operational use}
KBF should be used as a statistical triage tool, not as standalone proof of provider intent or backend identity.
A rejected audit means that the suspect endpoint is statistically inconsistent with the reference endpoint on the enrolled probe set under the audited interface.
Provider logs and independent route confirmation can support further investigation of the served backend and routing policy.
Repeated audits under fresh private probes can assess whether an inconsistency persists.
Auditors should keep the probe pool private and rotate probes across repeated audits.

\partitle{Future work}
KBF audits whether an endpoint serves the claimed model.
Future work should extend this view in two directions. First, high-stakes operational use requires stronger statistical calibration through repeated same-endpoint audits across days, accounts, provider routes, regions, and fresh private probe sets.
Second, auditors should measure beyond model identity, including routing transparency. A relay may fall back, mix providers, or apply user-specific routing policies while still advertising a single model.
Message integrity is another direction. Recent work shows that malicious intermediaries can inject tool payloads and observe secrets in transit~\cite{liu2026your}, and auditors need black-box tests for request rewriting, response mutation, and credential exposure.
A complete model-serving audit should also cover billing transparency, including inflated token counts, hidden cache hits, batch-price arbitrage, and account or free-credit resale.

\par\endgroup

\end{document}